\documentclass[journal, 10pt]{IEEEtran}
\usepackage{amsmath}
\usepackage{amssymb}
\usepackage{amsfonts}
\makeatletter

\newcommand{\Rmnum}[1]{\expandafter\@slowromancap\romannumeral #1@}
\makeatother
\usepackage{graphicx}
\usepackage{subfigure}
\usepackage{epsfig}
\usepackage{subfigure}
\usepackage{psfrag}
\usepackage{citesort}
\usepackage{xcolor}
\usepackage{enumerate}
\usepackage{extarrows}
\usepackage{balance}
\begin{document}
\title{To Harvest and Jam: A Paradigm of Self-Sustaining Friendly Jammers for Secure AF Relaying}
\author{Hong Xing, Kai-Kit Wong, Zheng Chu, and Arumugam Nallanathan
\thanks{This paper has been presented in part at the IEEE Global Communications Conference (GLOBECOM), Austin, TX, USA, December 8-12, 2014.}
\thanks{H. Xing and A. Nallanathan are with the Centre for Telecommunications Research, King's College London (e-mails: $\rm hong.xing@kcl.ac.uk$; $\rm arumugam.nallanathan@kcl.ac.uk$).}
\thanks{K.-K. Wong is with the Department of Electronic and Electrical Engineering, University College London (e-mail: $\rm kai$-$\rm kit.wong@ucl.ac.uk$).}
\thanks{Z. Chu is with the School of Electrical and Electronic Engineering, Newcastle University (e-mail: $\rm z.chu@ncl.ac.uk$).}}
\maketitle

\begin{abstract}
This paper studies the use of multi-antenna {\em harvest-and-jam (HJ)} helpers in a multi-antenna amplify-and-forward (AF) relay wiretap channel assuming that the direct link between the source and destination is broken. Our objective is to maximize the secrecy rate at the destination subject to the transmit power constraints of the AF relay and the HJ helpers. \textcolor{black}{In the case of perfect channel state information (CSI), the joint optimization of the artificial noise (AN) covariance matrix for cooperative jamming and the AF beamforming matrix is studied using semi-definite relaxation (SDR) which is tight, while suboptimal solutions are also devised with lower complexity. For the imperfect CSI case, we provide the equivalent reformulation of the worst-case robust optimization to maximize the minimum achievable secrecy rate. Inspired by the optimal solution to the case of perfect CSI, a suboptimal robust scheme is proposed striking a good tradeoff between complexity and performance.} Finally, numerical results for various settings are provided to evaluate the proposed schemes.
\end{abstract}

\begin{IEEEkeywords}
Harvest-and-jam, artificial noise, cooperative jamming, amplify-and-forward relay, wireless energy transfer, physical-layer security, semi-definite relaxation, robust optimization.
\end{IEEEkeywords}

\IEEEpeerreviewmaketitle
\setlength{\baselineskip}{1\baselineskip}
\newtheorem{definition}{\underline{Definition}}[section]
\newtheorem{fact}{Fact}
\newtheorem{assumption}{Assumption}
\newtheorem{theorem}{\underline{Theorem}}[section]
\newtheorem{lemma}{\underline{Lemma}}[section]
\newtheorem{corollary}{\underline{Corollary}}[section]
\newtheorem{proposition}{\underline{Proposition}}[section]
\newtheorem{example}{\underline{Example}}[section]
\newtheorem{remark}{\underline{Remark}}[section]
\newtheorem{algorithm}{\underline{Algorithm}}[section]
\newcommand{\mv}[1]{\mbox{\boldmath{$ #1 $}}}

\section{Introduction}\label{sec:Introduction}
The pressing demand for high data rate in wireless communications networks coupled with the fact that mobile devices are physically small and power-limited by batteries, has driven the notion of energy harvesting (EH) to become a promising resolution for green communications \cite{varshney2008transporting,Grover_Tesla_C10}. Among the varied available resources for EH, radio-frequency (RF)-enabled wireless energy transfer (WET) has aroused an upsurge of interest for its long operation range, ubiquitous existence in the electromagnetic radiation, and effective  energy multicasting, which motivates the paradigm of simultaneous wireless information and power transfer (SWIPT), e.g.,  \cite{Rui_TWC_SWIPT_J13,ZhouXun_WIPT_J13,Liu2013opportunistic,Xu2013Multiuser}.

A typical SWIPT system consists of one access point (AP) that has constant power supply and broadcasts wireless signals to a group of user terminals, amongst which some intend to decode information, referred to as information receivers (IRs), while others scavenge energy from the ambient radio signals, named energy receivers (ERs). This gives rise to a challenging physical (PHY)-layer security issue where the ERs may eavesdrop the information sent to the IRs due to their close proximity to the AP To overcome this problem, in \cite{Liu2014Secrecy,Ng2014Robust,xing2014secrecySWIPT}, several researchers presented various approaches to guarantee secret communication to the IRs and maximize the energy simultaneously transferred to the ERs or to satisfy the individual EH requirement for the ERs and maximize the secrecy rate for the IR, by advocating the dual use of the artificial noise (AN) or jamming.

However, previous works all assumed that the ERs in the SWIPT systems attempt to intercept the information for the IR, which is overly protective. On the contrary, it is possible that some ERs are cooperative, especially when they are EH-enabled wirelessly. Following the recent advances in \emph{wireless powered communications networks} \cite{ju2014throughput,xing2014harvest-and-jam}, this paper proposes a self-sustaining {\em harvest-and-jam (HJ)} relaying protocol, where in the first transmission phase a single-antenna transmitter transfers confidential information to a multiple-antenna amplify-and-forward (AF) relay and power to a group of multi-antenna EH-enabled idle helpers simultaneously, while in the second phase, the relay amplifies and forwards the information to the IR under the protection of the AN generated by the helpers using the energy harvested from their received signals in the first transmission phase.

Physical (PHY)-layer security issues in the rapidly growing cooperative networks have attracted much attention. Cooperative approaches, such as, \emph{cooperative jamming}, communications have been widely examined \cite{tekin2008general,dong2010improving,tang2008gaussian,Huang2011CJ}. The idea is to assist the transmitter in the secrecy transmission by generating an AN to interfere with the eavesdropper via either multiple antennas or external trusted helpers \cite{Goel2008,Zheng2011CJ,chu2014secrecy,cumanan2014secrecy}. However, all of those utilizing ANs require additional supply of power and therefore incur extra system costs. Meanwhile, collaborative use of relays to form effective beams jamming the eavesdropper, i.e., \emph{secure collaborative relay beamforming}, has been studied for relay-wiretap channels with single eavesdropper in \cite{zhang2010relay}, multiple eavesdroppers with AF relays and decode-and-forward (DF) relays in \cite{yang2013cooperative} and \cite{jiangyuan2011CRB}, respectively. All, however, assumed the availability of perfect channel state information (CSI). Though \cite{wang2013secure} proposed  robust AF relay beamforming against the eavesdropper's channel, the solutions were yet suboptimal.

The assumption of perfect CSI of the eavesdroppers appears to be too ideal because the eavesdroppers, despite being legitimate users, wish to hide from the transmitter without being cooperative in the stage of channel estimation. Even if they are registered users and  bound to help the transmitter in obtaining their CSIs to facilitate their own communication, the CSIs at the transmitter side will change due to mobility and Doppler effect, and may be outdated. Moreover, even for the legitimate users, the estimated CSIs may also be subject to quantization errors due to the limited capacity of the feedback channel, although the inaccuracy is reasonably assumed less severe than that for the eavesdroppers. To tackle this issue, state-of-art schemes have been developed (\cite{he2013wireless} and the references therein), among which the \emph{worst-case secrecy rate} is commonly employed to formulate the robust secrecy rate maximization problem \cite{li2011optimal,swindlehurst2012robust,Ng2014Robust,cumanan2014secrecy,chu2015robust}. The robust transmit covariance design for the secrecy rate maximization in a multiple-input-single-output (MISO) channel overheard by multi-antenna eavesdroppers was considered in \cite{li2011optimal,chu2015MISOME} while the enhanced secrecy performance was achieved by introducing a friendly jammer in the same scenario in \cite{swindlehurst2012robust}, in which a joint optimization of the robust transmit covariance and power allocation between the source and the helper was studied via geometric programming. More recently, \cite{Ng2014Robust} studied a joint robust design of the information beams, the AN and the energy signals for SWIPT networks with quality-of-service (QoS) constraints.

\textcolor{black}{The contribution of this paper is threefold. First, with perfect CSI, in addition to the joint optimal solutions, we propose two near-optimal schemes with much reduced complexity by exploiting the optimal structure of the relay weight matrix, and providing a semi-closed form solution for the relay weight matrix given fixed \emph{null-space} jamming, respectively. Second, besides the imperfect eavesdropper's channel, legitimate channels such as those from the $K$ HJ helpers (the transmitter) to the legitimate receiver ($K$ HJ helpers), and from the AF relay to the receiver are jointly modeled with imperfect estimation, and multiple semi-indefinite non-convex constraints have been judiciously replaced by linear matrix inequalities (LMIs) to fit the semi-definite programming (SDP). Third, a rank-one reconstruction algorithm exploiting the structure of the semi-definite relaxation (SDR)-based solutions has been proposed to provide promising performance at low computational cost.}

Of particular relevance to our work is \cite{li2015robust} which jointly optimizes the AF matrices and AN covariances in a relay wiretap channel with multiple multi-antenna AF relays and multiple multi-antenna eavesdroppers via a worst-case robust formulation. While our network model is similar, the difference of our work from \cite{li2015robust} is twofold. On one hand, in this paper, the AN generated by the friendly jammers are subject to their respective channels from the transmitter during WET in the first transmission phase. On the other hand, the technique in \cite[\emph{Proposition 1}]{li2015robust} cannot be applied to our problem since the AN beams and the forwarded information are transmitted via different channels in ours. As a consequence, to the best of authors' knowledge, our proposed worst-case based {\em robust} optimization scheme that incorporates imperfect CSIs into all the HJ helpers, has not been addressed in the literature.

It is worth noting that devising a wireless-powered friendly jammer to enhance PHY-layer security for a direct transmission protocol was studied in \cite{xiangyun2014secure}, in which the ``harvesting'' blocks and ``jamming'' blocks were well exploited to compose four different types of harvesting-jamming cycles. Compared to \cite{xiangyun2014secure}, which focused on the dedicated scheduling of ``harvest'' and ``jam'' operations and its long-term performance, ours are concerned with adaptive rate/power optimization with multiple HJ helpers to achieve higher worst-case secrecy rate. \textcolor{black}{Moreover, instead of assuming perfect channels to/from the HJ helpers, our {\em robust} optimization algorithm takes imperfect legitimate channels into account to provide robustness.}

Note that in this paper, as in \cite{wang2013secure,li2015robust}, we assume that the channel between the transmitter and the AF relay is perfectly known and there is no direct link between the transmitter and the receiver or the eavesdropper, a common assumption in the concerned AF relay wiretap channel \cite{zhang2010relay,yang2013cooperative}.

{\em Notations}---Throughout, we use the upper case boldface letters for matrices and lower case boldface letters for vectors. The superscripts $(\cdot)^{T}$, $(\cdot)^{\dagger}$ and $(\cdot)^{H}$ represent the transpose, conjugate and conjugate transpose, respectively. Also, ${\rm tr}(\cdot)$ and $\mathbb{E}[\cdot]$ stand for the trace of a matrix and the statistical expectation for random variables, respectively. Likewise, ${\rm vec}(\mv A)$ is defined as a column vector obtained by stacking the rows of \(\mv A\) on top of one another. ${\rm vec}^{(-1)}$ is the inverse operation of ${\rm vec}$. $\mathbf{null}(\mv A)$ denotes the null space of \(\mv A\). $\otimes$ represents the Kronecker product of two matrices. In addition, the notation $\mv{A}\succeq 0$ indicates that $\mv{A}$ is a positive semi-definite matrix and $\mv{I}$ ($\mv 0$) denotes an identity (all-zero) matrix with appropriate size. Furthermore, $\|\cdot\|$ represents the Euclidean norm of a vector, while $P_r(\cdot)$ stands for the probability of an input random event. Finally, $[x]^{+}$ denotes $\max(0,x)$ and $(\cdot)^\ast$ stands for an optimal solution.

\section{Network Model}\label{sec:System Model}
\begin{figure}[htb]
\centering
\epsfxsize=1\linewidth
\subfigure[AF-relaying wiretap channel with jamming.]{\label{fig:subfig:channel model}\includegraphics[width=7.0cm]{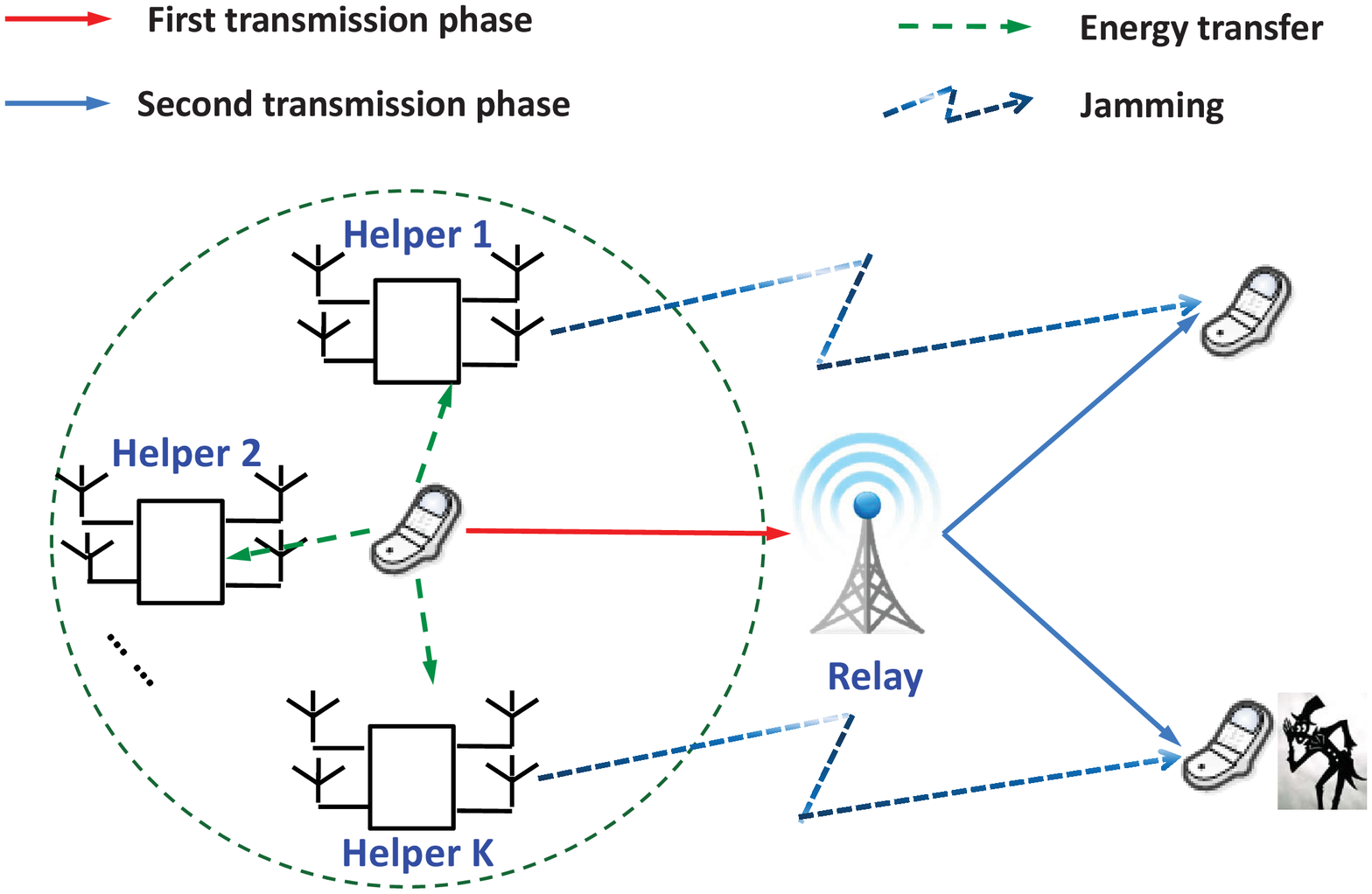}}
\vfill
\epsfxsize=1\linewidth
\subfigure[\textcolor{black}{The {\em HJ} relaying protocol.}]
{\label{fig:subfig:HJ protocol}\includegraphics[width=7.0cm]{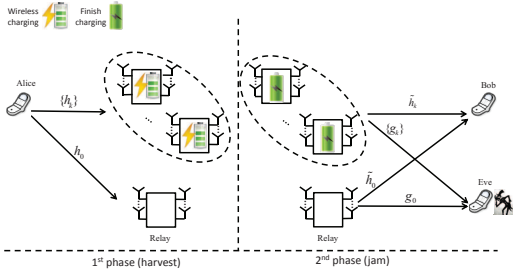}}
\caption{\textcolor{black}{HJ-enabled cooperative relaying for secure SWIPT.}}\label{fig:system model}
\vspace{-1.5em}
\end{figure}

We consider a cooperative relay wiretap channel for SWIPT over a given frequency band as shown in Fig.~\ref{fig:subfig:channel model}. We assume that there is a transmitter, named Alice, sending confidential messages to the IR, Bob, in the presence of an eavesdropper \cite{chorti2013resilience}, Eve, with the aid of a multi-antenna AF relay and \(K\) ERs willing to act as HJ helpers, \(\mathcal{H}_{\rm helper}=\{{\sf H}_1,\dots,{\sf H}_K\}\).  \textcolor{black}{The transmitter, ERs, and the AF relay are deployed in a same cluster that is relatively far away from the destination and Eve, such that there is no direct link from the transmitter to the receiver or Eve, respectively. Moreover, the ERs are assumed to be located closer to the transmitter than the AF relay in order that they can harvest sufficient amount of energy for jamming.} Alice, Bob and Eve are all assumed to be equipped with single antenna, while the AF relay and each of the \(K\) helpers are assumed to have the same \(N_t\) antennas.

Using two equal slots for the HJ relaying protocol, as shown in \textcolor{black}{Fig.~\ref{fig:subfig:HJ protocol}}, for the first phase, Alice sends a confidential message to the relay while simultaneously transferring energy to the \(K\) helpers; for the second phase, the relay amplifies and forwards the message to Bob while the \(K\) helpers perform cooperative jamming using their respective harvested energy from the first transmission phase, to compromise Eve. In this paper, we assume a quasi-static fading environment and for convenience denote \(\mv h_0\in\mathbb{C}^{N_t\times 1}\) as the complex channel from the transmitter to the relay and \(\mv h_k\in\mathbb{C}^{N_t\times 1}\), \(k=1,\ldots,K\), as that from the transmitter to the \(k\)th helper; \(\tilde{\mv h}_0\) as the transpose of the complex channel from the relay to Bob and \(\tilde{\mv h}_k\in\mathbb{C}^{N_t\times1}\), \(k=1,\ldots,K\),  as that from \({\sf H}_k\) to Bob; \(\mv g_0\in\mathbb{C}^{N_t\times 1}\) and \(\mv g_k\in\mathbb{C}^{N_t\times 1}\), \(k=1,\ldots,K\), as those from the relay and \({\sf H}_k\) to Eve, respectively.

In the first transmission phase, the baseband received signal at the AF relay can be expressed as
\begin{equation}
  \mv{y}_r=\mv{h}_0\sqrt{P_s}s+\mv{n}_r , \label{eq:received at the relay}
\end{equation}
where \(s\) is a circularly symmetric complex Gaussian (CSCG) random variable, denoted by \(s \sim \mathcal{CN}(0,1)\) and \(\mv n_r\) is the additive complex noise vector, denoted by \(\mv{n}_r \sim \mathcal{CN}(\mv{0}, \sigma^2_r\mv{I})\). Also, \(P_s\) denotes the given transmit power at Alice. Further, the received signal at each helper \({\sf H}_k\) is expressed as
\begin{equation}\label{eq:received at the CJ helpers}
   \mv{y}_k=\mv{h}_k\sqrt{P_s}s+\mv{n}^\prime_k ,
\end{equation}
where \(\mv{n}^\prime_k\) is the additive noise, denoted by \(\mv{n}^\prime_k\sim\mathcal{CN}(\mv{0},\sigma_h^2\mv{I})\).

On the other hand, for WET, the harvested energy 
of \({\sf H}_k\) in each unit slot is given by
\begin{equation}\label{eq:harvested at HJ helpers}
  E_k=\eta \mathbb{E}[\| \mv{h}_k\sqrt{P_s}s\|^2]=\eta P_s\left\|\mv{h}_k\right\|^2,\; \forall k,
\end{equation}
where \(0<\eta\leq 1\) denotes the EH efficiency.

In the second transmission phase, the linear operation at the AF relay can be represented by
\begin{equation}\label{eq:transmitted signal at the relay}
 \mv x^\prime=\mv{Wy}_r,
\end{equation}
where \(\mv x^\prime\in\mathbb{C}^{N_t\times1}\) is the retransmit signal at the AF relay and \(\mv W\in\mathbb{C}^{N_t\times N_t}\) is the beamforming matrix. Note that the transmit power of the AF relay can be shown as
\begin{equation}\label{eq:transmit power at the relay}
{\rm tr}\left(\mathbb{E}\left[\mv x\mv x^H\right]\right)={\rm tr}\left(\mv W\left(P_s\mv{h}_0\mv{h}^H_0  +\sigma^2_r\mv{I}\right)\mv W^H \right),
\end{equation}
which is constrained by the maximum available power at the AF relay, i.e., \(P_r\), which is given by
\begin{equation}\label{eq:transmit power constaint at the relay}
{\rm tr}\left(\mv W\left(P_s\mv{h}_0\mv{h}^H_0  +\sigma^2_r\mv{I}\right)\mv W^H \right )\leq P_r.
\end{equation}
In the meantime, each \({\sf H}_k\) will help generate an AN \(\mv n_k\in\mathbb{C}^{N_t\times1}\) to interfere with Eve. Similar to \cite{Goel2008}, we assume that \(\mv n_k\)'s are independent CSCG vectors denoted by \(\mv n_k\sim\mathcal{CN}(0,\mv Q_k)\), \(\forall k\), since the worst-case noise for   Eve is known to be Gaussian. In addition, each \({\sf H}_k\) has a transmit power constraint due to its harvested energy in the previous transmission phase, i.e., \({\rm tr}(\mv Q_k)\le\eta P_s\|\mv h_k\|^2\) (c.f.~\eqref{eq:harvested at HJ helpers}), \(\forall k\).

The received signal at Bob can thus be expressed as
\begin{equation}\label{eq:received at Bob}
\mv{y}_b 
=\sqrt{P_s}\widetilde{\mv{h}}^T_0\mv{Wh}_0s+\sum^K_{k=1}\widetilde{\mv{h}}^T_k\mv{n}_k+\widetilde{\mv{h}}_0^T\mv{Wn}_r+\mv{n}_b,
\end{equation}
where \(\mv n_b\sim\mathcal{CN}(0,\sigma_b^2\mv I)\) is the additive noise at Bob. Similarly, the received signal at Eve can be expressed as
\begin{align}
  \mathbf{y}_e=\sqrt{P_s}\mv{g}^T_0\mv{Wh}_0s +\sum^K_{k=1}\mv{g}^T_k\mv{n}_k+\mv{g}_0^T\mv{Wn}_r+\mv{n}_e, \label{eq:received at Eve}
\end{align}
where \(\mv n_e\sim\mathcal{CN}(0,\sigma_e^2\mv I)\). According to \eqref{eq:received at Bob} and \eqref{eq:received at Eve}, the signal-to-interference-plus-nose-ratio (SINR) at Bob and Eve can be, respectively, expressed as
\begin{equation}\label{eq:SINR at Bob}
  \gamma_b=\frac{P_s\vert\widetilde{\mv{h}}^T_0\mv{Wh}_0\vert^2}{\sigma^2_r\widetilde{\mv{h}}^T_0\mv{WW}^H\widetilde{\mv{h}}_0^\dagger+\sum^K_{k=1}\widetilde{\mv{h}}^T_k\mv{Q}_k\widetilde{\mv{h}}_k^\dagger+\sigma^2_b},
\end{equation}
and
\begin{equation}\label{eq:SINR at Eve}
 \gamma_e=\frac{P_s\vert\mv{g}^T_0\mv{Wh}_0\vert^2}{\sigma^2_r\mv{g}^T_0\mv{WW}^H\mv{g}_0^\dagger+\sum^K_{k=1}\mv{g}^T_k\mv{Q}_k\mv{g}_k^\dagger+\sigma^2_e}.
\end{equation}
As such, the achievable secrecy rate at Bob is \cite{Goel2008}
\begin{equation}\label{eq:achievable secrecy rate}
 r_0=\frac{1}{2}\left[\log_2(1+\gamma_b)-\log_2(1+\gamma_e)\right ]^+.
\end{equation}

\section{Joint AN-AF Beamforming with Perfect CSI}\label{sec:A Joint Optimization Based on Perfect CSI}
\subsection{Problem Formulation for Perfect CSI}\label{subsec:Problem Formulation for perfect CSI}
We aim to maximize the secrecy rate at Bob subject to the transmit power constraints at the AF relay and each individual helper \({\sf H}_k\), \(k=1,\dots,K\). Thus, our problem is to solve
\begin{subequations}
 \begin{align}\mathrm{(P1)}:~\mathop{\mathtt{max}}_{\{\boldsymbol{Q}_k\},\boldsymbol{W}}
& ~~~ r_0\notag \\
\mathtt {s.t.}& ~~~\eqref{eq:transmit power constaint at the relay},\label{eq:power constraint at the relay}\\
& ~~~{\rm tr}\left(\mv{Q}_k \right )\leq \eta P_s\left\|\mv{h}_k\right\|^2, \; \forall k,\label{eq:power constraint for the AN}\\
& ~~~\mv{Q}_k\succeq \mv 0, \; \forall k\label{eq:constraint on PSD}.
\end{align}
\end{subequations}
Next, we define a new function \(\bar F(\{\mv{Q}_k\},\mv W)\) as
\begin{equation}\label{eq:def of bar F}
\bar F(\{\mv{Q}_k\},\mv W)\triangleq\frac{1+\gamma_b}{1+\gamma_e}.
\end{equation}
It can be easily shown that the optimal solution \(\{\mv Q_k^\ast\}\), \(\mv W^\ast\) to $({\rm P1})$, is also optimal for $({\rm P1^\prime})$ given by
\begin{equation}\mathrm{(P1^\prime)}:~\mathop{\mathtt{max}}_{\{\boldsymbol{Q}_k\},\boldsymbol{W}}
\bar F(\{\mv{Q}_k\},\mv W)
~\mathtt {s.t.} ~\eqref{eq:power constraint at the relay}-\eqref{eq:constraint on PSD}.
\end{equation}

Hence, we focus on solving problem $({\rm P1}^\prime)$ in the rest of the paper. However, since $({\rm P1}^\prime)$ is in general a non-convex problem that is hard to solve, we will reformulate it into a two-stage optimization problem. First, we constrain the SINR at Eve to be \(\bar\gamma_e\), it thus follows from \eqref{eq:def of bar F} that \(\bar F(\{\overline{\mv Q}_k\},\mv W)\) is maximized when \(\gamma_b\) is maximized, which can be obtained by solving the following problem:
\begin{multline}\label{eq:SNR constaint at the Eve}
\mathrm{(P1^\prime.1)}: \mathop{\mathtt{max}}_{\{\boldsymbol{Q}_k,\boldsymbol{W}\}}
\frac{P_s\vert\widetilde{\mv h}_0^T{\mv W}\mv h_0\vert^2}{\sigma_r^2\widetilde{\mv h}_0^T{\mv W}{\mv W}^H\widetilde{\mv h}_0^\dagger+\sum_{k=1}^K\tilde{\mv h}_k^T\mv Q_k\tilde{\mv h}_k^\dag+\sigma_b^2}\\
\mathtt{s.t.}~ \frac{P_s\vert\mv g_0^T{\mv W}\mv h_0\vert^2}{\sigma_r^2\mv g_0^T{\mv W}{\mv W}^H\mv g_0^\dagger+\sum_{k=1}^K{\mv g}_k^T\mv Q_k{\mv g}_k^\dag+\sigma_e^2}\!=\!\bar\gamma_e,\\
\eqref{eq:power constraint at the relay}-\eqref{eq:constraint on PSD}.
\end{multline}
Let \(H(\bar\gamma_e)\) denote the optimal value of $({\rm P1}^\prime.1)$ given \(\bar\gamma_e\). Then $({\rm P1}^\prime)$ can be equivalently solved by
\begin{equation}
\mathrm{(P1^\prime.2)}: \mathop{\mathtt{max}}_{\bar\gamma_e>0}~\frac{1+H(\bar\gamma_e)}{1+\bar\gamma_e}. \label{eq:(P1'.2)}
\end{equation}


\begin{lemma}\label{lemma:P1'.2 same optimal value}
Problem $({\rm P1}^\prime)$ has the same optimal value as $({\rm P1}^\prime.2)$, and the same optimal solution as $({\rm P1}^\prime.1)$ when \(\bar\gamma_e\) takes the optimal solution for $({\rm P1}^\prime.2)$.
\end{lemma}

\begin{IEEEproof}
The proof follows from \cite[\emph{Lemmas 4.1-4.2}]{Liu2014Secrecy}.
\end{IEEEproof}

Therefore, $({\rm P1}^\prime)$ can be solved in the following two steps. First, given any \(\bar\gamma_e>0\), we solve $({\rm P1}^\prime.1)$ to attain \(H(\bar\gamma_e)\); then we solve $({\rm P1}^\prime.2)$ to obtain the optimal \(\bar\gamma_e^\ast\). 
\vspace{-1.0em}
\subsection{Optimal Solution to $({\rm P1}^\prime.1)$}\label{subsec:Optimal Solution}
Here, we consider solving problem $(\rm P1^\prime.1)$ by jointly optimizing the covariance matrix for the AN at each of the HJ helper, \(\mv Q_k\)'s, and the beamforming matrix, \(\mv W\). To facilitate the analysis in the sequel, we rewrite the following equations in line with our definition of \({\rm vec}(\cdot)\) \cite[Chapter 13]{laub2005matrix}:
\begin{align}
\vert\tilde{\mv h}_0^T\mv W\mv h_0\vert^2 & =\vert{\rm vec}^T(\tilde{\mv h}_0\mv h_0^T){\rm vec}(\mv W)\vert^2, \\
\tilde{\mv h}_0^T\mv W\mv W^H\tilde{\mv h}_0^\dagger & =\|\tilde{\mv h}_0^T\otimes\mv I{\rm vec}(\mv W)\|^2,\\
\vert\mv g_0^T\mv W\mv h_0\vert^2 & =\vert{\rm vec}^T(\mv g_0\mv h_0^T){\rm vec}(\mv W)\vert^2,\\
\mv g_0^T\mv W\mv W^H\mv g_0^\dagger & =\|\mv g_0^T\otimes\mv I{\rm vec}(\mv W)\|^2.
\end{align}
In addition, \({\rm tr}(\mv W(P_s\mv h_0\mv h_0^H+\sigma_r^2\mv I)\mv W^H)=\|\mv \Phi\mv w\|^2\), where \(\mv\Phi=(\mv I\otimes\mv\Theta^T)^{1/2}\) with \(\mv\Theta=P_s\mv h_0\mv h_0^H+\sigma_r^2\mv I\). Hence, $\rm (P1^\prime.1)$ can be rewritten as
\begin{subequations}
\begin{align}
\mathrm{(P1^\prime.1\text{-}RW)}:\nonumber\\ \mathop{\mathtt{max}}_{\boldsymbol{W},\{\boldsymbol{Q}_k\}}
&~~~\frac{P_s\vert\mv f_1^T\mv w\vert^2}{\sigma_r^2\left\|\mv Y_1\mv w\right\|^2+\sum_{k=1}^K\tilde{\mv h}_k^T\mv Q_k\tilde{\mv h}_k^\dagger+\sigma_b^2}\notag\\
\mathtt{s.t.}& ~~~\frac{P_s\vert\mv f_2^T\mv w\vert^2}{\sigma_r^2\left\|\mv Y_2\mv w\right\|^2+\sum_{k=1}^K\mv g_k^T\mv Q_k\mv g_k^\dagger+\sigma_e^2}=\bar \gamma_e,\\
& ~~~\left\|\Phi \mv w\right\|^2\leq P_r,\\
& ~~~\eqref{eq:power constraint for the AN}, \eqref{eq:constraint on PSD}\notag,
\end{align}
\end{subequations}
in which \(\mv w={\rm vec}(\mv W)\), \(\mv f_1={\rm vec}(\widetilde{\mv h}_0\mv h_0^T)\), \(\mv f_2={\rm vec}(\mv g_0\mv h_0^T)\), \(\mv Y_1=\widetilde{\mv h}_0^T\otimes \mv I\) and \(\mv Y_2=\mv g_0^T\otimes\mv I\).

As problem $\rm (P1^\prime.1\text{-}RW)$ is non-convex, we define \(\mv X\triangleq\mv w\mv w^H\), \(\mv F_1\triangleq\mv f_1^\dagger\mv f_1^T\), \(\mv F_2\triangleq\mv f_2^\dagger\mv f_2^T\), \(\overline{\mv Y}_1\triangleq\mv Y_1^H\mv Y_1 \), \(\overline{\mv Y}_2\triangleq\mv Y_2^H\mv Y_2 \) and \(\overline{\mv \Phi}\triangleq\mv \Phi^H\mv\Phi\). Then by ignoring the rank-one constraint on \(\mv X\), $\rm (P1^\prime.1\text{-}RW)$ is modified as
\begin{subequations}\label{eq:constraints for P1'.1-RW-SDR-Eqv}
\begin{align}
&\!\!\!\!\!\!\!\!\!\!\!\!\!\!\!\!\!\!\!\!\!\!\!\!\!\!\!\!\!\!\!\!\!\!~~~~~ ~~~ \mathrm{(P1^\prime.1\text{-}RW\text{-}SDR\text{-}Eqv)}:\nonumber\\ \!\!\!\!\!\mathop{\mathtt{max}}_{\boldsymbol{X},\{\boldsymbol{Q}_k\}}
 &~~~\frac{P_s{\rm tr}(\mv F_1\mv X)}{\sigma^2_r{\rm tr}(\overline {\mv Y}_1\mv X)+\sum_{k=1}^K\tilde{\mv h}_k^T\mv Q_k\tilde{\mv h}_k^\dagger+\sigma_b^2}\nonumber\\
 \mathtt{s.t.}& ~~~P_s{\rm tr}(\mv F_2\mv X)\nonumber\\
&~~~ =\bar\gamma_e\left(\sigma^2_r{\rm tr}(\overline {\mv Y}_2\mv X)+\sum_{k=1}^K\mv g_k^T\mv Q_k\mv g_k^\dagger+\sigma_e^2\right), \\
&~~~{\rm tr}(\overline{\mv \Phi}\mv X)\leq P_r,\\
&~~~\mathrm{tr}\left(\mv{Q}_k \right)\leq \eta P_s\left\|\mv{h}_k\right\|^2, \; \forall k,\\
&~~~\mv X\succeq \mv 0, \, \mv Q_k\succeq \mv 0, \; \forall k.
\end{align}
\end{subequations}
Problem $\rm (P1^\prime.1\text{-}RW\text{-}SDR\text{-}Eqv)$, via Charnes-Cooper transformation \cite{charnes1962programming}, can be equivalently recast as
\begin{subequations}\label{eq:transformation by C-O}
\begin{align}
&\mathrm{(P1^\prime.1\text{-}RW\text{-}SDR)}:~\mathop{\mathtt{max}}_{\boldsymbol{X},\{\boldsymbol{Q}_k\},\tau}
~~~P_s{\rm tr}(\mv F_1\mv X)\nonumber\\
&\mathtt{s.t.}~~~\sigma^2_r{\rm tr}(\overline {\mv Y}_1\mv X)+\sum_{k=1}^K\tilde{\mv h}_k^T\mv Q_k\tilde{\mv h}_k^\dagger+\tau\sigma_b^2=1, \label{eq:constraint on C-O transformation}\\
&P_s{\rm tr}(\mv F_2\mv X)\notag\\
&=\bar\gamma_e\left(\sigma^2_r{\rm tr}(\overline {\mv Y}_2\mv X)+\sum_{k=1}^K\mv g_k^T\mv Q_k\mv g_k^\dagger+\tau\sigma_e^2\right), \label{eq:constraint on SINR of Eve}\\
&{\rm tr}(\overline{\mv \Phi}\mv X)\leq \tau P_r, \label{eq:constraint on transmit power of the relay}\\
&\mathrm{tr}\left(\mv{Q}_k \right)\leq \tau\eta P_s\left\|\mv{h}_k\right\|^2, \; \forall k, \label{eq:constraint on amount of AN for each HJ helper}\\
&\mv X\succeq \mv 0, \, \mv Q_k\succeq \mv 0, \; \forall k, \, \tau\ge 0. \label{eq:constraint on X, Qk and tau}
\end{align}
\end{subequations}

\begin{lemma}\label{lemma:inequality equivalent of (P1.1-RW-SDR)}
The constraints in \eqref{eq:constraint on C-O transformation} and \eqref{eq:constraint on SINR of Eve} can be replaced by \(\sigma^2_r{\rm tr}(\overline {\mv Y}_1\mv X)+\sum_{k=1}^K\tilde{\mv h}_k^T\mv Q_k\tilde{\mv h}_k^\dagger+\tau\sigma_b^2\le1\) and \(P_s{\rm tr}(\mv F_2\mv X)\le\bar\gamma_e(\sigma^2_r{\rm tr}(\overline {\mv Y}_2\mv X)+\sum_{k=1}^K\mv g_k^T\mv Q_k\mv g_k^\dagger+\tau\sigma_e^2)\), respectively, where both inequalities will be activated when problem $\rm (P1^\prime)$ obtains its optimum value.
\end{lemma}

\begin{IEEEproof}
See \cite[Appendix \ref{appendix:proof of prop:structure of optimal X and its rank-one reconstruction}]{xing2015robust}.
\end{IEEEproof}

Since problem $\rm (P1^\prime.1\text{-}RW\text{-}SDR)$ is a standard convex optimization problem and satisfies the Slater's condition, its gap with its dual problem is zero \cite{boyd2004convex}. Now, let \(\lambda\) denote the dual variable associated with the equality constraint in \eqref{eq:constraint on C-O transformation}, \(\alpha\) associated with the other equality constraint in \eqref{eq:constraint on SINR of Eve}, \(\beta_0\) associated with the transmit power constraint for the AF relay in \eqref{eq:constraint on transmit power of the relay}, \(\{\beta_k\}\) associated with the transmit power constraints for each \({\sf H}_k\) in \eqref{eq:constraint on amount of AN for each HJ helper}, and \(\zeta\) associated with \(\tau\). Then the Lagrangian of problem $\rm (P1^\prime.1\text{-}RW\text{-}SDR)$ is given by
\begin{eqnarray}\label{eq:Lagrangian of (P1.1-RW-SDR)}
L(\mv\Omega)&\!\!=&\!\!\mathrm{tr}(\mv A\mv X)+\sum_{k=1}^K\mathrm{tr}(\mv B_k\mv Q_k)+\zeta\tau+\lambda,
\end{eqnarray}
where \(\mv\Omega\) denotes the set of all primal and dual variables,
\begin{align}
\mv A&=P_s\mv F_1-\lambda\sigma_r^2\overline{\mv Y}_1-\alpha P_s\mv F_2+\alpha\bar\gamma_e\sigma_r^2\overline{\mv Y}_2-\beta_0\overline{\mv \Phi}, \label{eq:A}\\
\mv B_{k}&=-\lambda\tilde{\mv h}_k^\ast\tilde{\mv h}_k^T+\alpha\bar\gamma_e\mv g_k^\ast\mv g_k^T-\beta_k\mv I,\; \forall k,  \label{eq:Bk}\\
\zeta&=-\lambda\sigma_b^2+\alpha\bar\gamma_e\sigma_e^2+\beta_0P_r+\sum_{k=1}^K\eta P_s\beta_k\|\mv h_k\|^2. \label{eq:zeta}
\end{align}

\begin{proposition}
The optimal solution, \((\mv X^\ast, \{\mv Q_k^\ast\}, \tau^\ast)\), to $\rm (P1^\prime.1\text{-}RW\text{-}SDR)$ satisfies the following conditions:
\begin{enumerate}
    \item 
    $\mathrm{rank}(\mv Q_k)\left\{\begin{array}{ll}\ge N_t-2,\ &{\rm if}\, \beta_k^\ast=0,\\
    =1, \ &{\rm if}\, \beta_k^\ast>0, \end{array}\right.\forall k$;
    \item \(\mv X^\ast\) can be expressed as
    \begin{align}
     \mv X^\ast=\sum_{n=1}^{N_t^2-r_c}a_n\mv \eta_n\mv \eta_n^H+b\mv \xi\mv \xi^H, \label{eq:structure of optimal X}
    \end{align}
    where \(a_n\ge 0\) \(\forall n\), \(b>0\), \(r_c=\mathrm{rank}(\mv C^\ast)\) (c.f. \eqref{eq:A star}) and \(\mv\xi\in\mathbb{C}^{N_t^2\times1}\) is a vector orthogonal to \(\mv\Xi=\{\mv\eta_n\}_{n=1}^{N_t^2-r_c}\), which consists of orthonormal basis for \(\mathbf{null}(\mv C^\ast)\);
    \item According to \eqref{eq:structure of optimal X}, if \(\mathrm{rank}(\mv X^\ast)>1\), then we have the following sufficient condition to yield an optimal solution of \(\mv X\) with rank-one:
        \begin{align}
       \hat{\mv X}^\ast& =b\mv \xi\mv \xi^H,\label{eq:reconstructed structure of optimal hat X}\\
       \hat{\mv Q}_k^\ast&=\mv Q_k^\ast,\; \forall k,\label{eq:reconstructed Qk}\\
       \hat \tau^\ast&=\tau^\ast+\Delta \tau, \label{eq:reconstructed tau}
        \end{align}
      is also optimal to problem $\rm (P1^\prime.1\text{-}RW\text{-}SDR)$, if there exists \(\Delta\tau\ge 0\) such that
        \begin{multline}\label{eq:sufficient condition for rank-one X}
\left[\sum_{n=1}^{N_t^2-r_c}a_n\mathrm{tr}\left(\mv\eta_n^H(\tfrac{\sigma_r^2\overline{\mv Y}_2}{\sigma_e^2}-\tfrac{P_s\mv F_2}{\gamma_e\sigma_e^2})\mv\eta_n\right)\right]^+\\
\le\Delta\tau \le\tfrac{\sigma_r^2}{\sigma_b^2}\sum_{n=1}^{N_t^2-r_c}a_n\mathrm{tr}\left(\mv\eta_n^H\overline{\mv Y}_1\mv\eta_n\right).
        \end{multline}

\end{enumerate} \label{prop:structure of optimal X and its rank-one reconstruction}
\end{proposition}

\begin{IEEEproof}
See Appendix \ref{appendix:proof of prop:structure of optimal X and its rank-one reconstruction}.
\end{IEEEproof}

Note from Proposition \ref{prop:structure of optimal X and its rank-one reconstruction} that if \(\mathrm{rank}(\mv X^\ast)\!\!=\!\!1\), then the optimal \(\mv w^\ast\) to $\rm (P1^\prime.1\text{-}RW)$ can be found directly from the eigenvalue decomposition (EVD) of \(\overline{\mv X}^\ast\), where \(\overline{\mv X}^\ast=\mv X^\ast/\tau^\ast\). Namely, the upper-bound optimum value obtained by solving $\rm (P1^\prime.1\text{-}RW\text{-}SDR)$ is tight in this case; otherwise, \((\mv X^\ast, \{\mv Q_k^\ast\}, \tau^\ast)\) only serves as an upper-bound solution.

Now, we show that this upper-bound is always achievable by a rank-one \(
\mv X^\ast\). When \(\mathrm{rank}(\mv X^\ast)>1\), firstly, we check whether the sufficient condition proposed in \eqref{eq:sufficient condition for rank-one X} is satisfied. If it is met, then a direct reconstruction of \((\hat{\mv X}^\ast, \{\hat{\mv Q}_k^\ast\}, \hat\tau^\ast)\) with \(\mathrm{rank}(\hat{\mv X}^\ast)=1\) follows according to \eqref{eq:reconstructed structure of optimal hat X}--\eqref{eq:reconstructed tau}; otherwise, assume that any optimal solution to problem $\rm (P1^\prime.1\text{-}RW\text{-}SDR)$ has no zero component, i.e., \((\mv X^\ast\neq\mv 0,\{\mv Q_k^\ast\neq\mv 0\},\tau^\ast\neq 0)\). In addition, the number of optimization variables and the number of shaping constraints are denoted by \(L\) and \(M\), respectively. Since \(L=K+2\) and \(M=K+3\) for $\rm (P1^\prime.1\text{-}RW\text{-}SDR)$, we have \(M\le L+2\) satisfied. Thus, according to \cite[\emph{Proposition 3.5}]{huang2010rank}, $\rm (P1^\prime.1\text{-}RW\text{-}SDR)$ has  an optimal solution of \(\hat{\mv X}^\ast\) that is rank-one. Also, the detailed rank reduction procedure based on an arbitrary-rank solution has been given in \cite[Algorithm 1]{huang2010rank}. Algorithm \ref{table:Algorithm I} for solving \(\mathrm{(P1^\prime)}\) is shown in Table \ref{table:Algorithm I}.
\begin{table}[htp]
\begin{center}
\vspace{0.025cm}
\caption{\textcolor{black}{\rm Algorithm I for \(\mathrm{(P1^\prime)}\)}} \label{table:Algorithm I}
\vspace{-0.05cm}
 \hrule
\vspace{0.2cm}
\begin{itemize}
\item {\bf Initialize} \(\bar\gamma_{e\_{\rm search}}=0:\alpha:\bar\gamma_{e\max}\) and $i=0$
\item {\bf Repeat}
\begin{itemize}
\item [1)] {\bf Set} $i=i+1$;
\item [2)]  Given \(\bar\gamma_e=\bar\gamma_{e\_{\rm search}}(i)\),\\
{\bf solve} \(\mathrm{(P1^\prime.1\text{-}RW\text{-}SDR)}\) and {\bf obtain} \(H(\bar\gamma_e^{(i)})\).
\end{itemize}
\item {\bf Until} \(i=L\), where  \(L=\lfloor{\tfrac{\bar\gamma_{e\max}}{\alpha}}\rfloor+1\) is the length of \(\bar\gamma_{e\_{\rm search}}\)
\item {\bf Set} \(\bar\gamma_e^\ast=\bar\gamma_{e\_{\rm search}}\left(\!\arg\max\limits_{i}\!\left\{\tfrac{1+H(\bar\gamma_e^{(i)})}{1+\bar\gamma_e^{(i)}}\right\}\right)\) for \(\mathrm{(P1^\prime.2)}\)
\item Given \(\bar\gamma_e^\ast\), {\bf solve} \(\mathrm{(P1^\prime.1\text{-}RW\text{-}SDR)}\) to obtain \((\mv X^\ast, \{\mv Q_k^\ast\}, \tau^\ast)\)\\
    {\bf if} \({\rm rank}(\mv X^\ast)=1\), {\bf apply} EVD on \(\mv X^\ast\) such that \(\mv X^\ast=\mv w^\ast\mv w^{\ast H}\);\\
    {\bf else if} the sufficient condition in \eqref{eq:sufficient condition for rank-one X} is satisfied,\\
    {\bf construct} \((\hat{\mv X^\ast},\{\hat{\mv Q}_k^\ast\}, \hat\tau^\ast)\) following \eqref{eq:reconstructed structure of optimal hat X}-\eqref{eq:reconstructed tau} and {\bf set} \(\mv w^\ast=\sqrt{b}\mv\xi\);\\
    ~~~~{\bf else} {\bf construct} \(\hat{\mv X}^\ast\) using the procedure in \cite[{\rm Algorithm 1}]{huang2010rank}.\\
    ~~~~{\bf end}\\
    {\bf end}
    \item {\bf Recover} \(\mv W^\ast={\rm vec}^{-1}(\mv w^\ast)\)
\end{itemize}
\vspace{0.2cm} \hrule 
\end{center}
\end{table}

\subsection{Suboptimal Solutions to $\rm (P1^\prime.1)$}\label{subsec:Suboptimal Solutions}
\subsubsection{Optimal Solution Structure based Scheme}\label{subsubsec:structure-based}
We propose a relay beamforming design for $\rm (P1^\prime.1)$ based on the optimal structure of \(\mv W\) \cite[\underline{\emph{Theorem}}~3.1]{zhang2009optimal}. First, define \(\mv H_1\triangleq[\tilde{\mv h}_0\ \mv g_0]\) and \(\mv H_2\triangleq[\mv h_0\ \mv g_0]\). Then express the truncated singular-value decomposition (SVD) of \(\mv H_1\) and \(\mv H_2\), respectively, as
\begin{align}
 \mv H_1&=\mv U_1\mv \Sigma_1\mv V_1^H,\label{eq:SVD of H1}\\
\mv H_2&=\mv U_2\mv \Sigma_2\mv V_2^H. \label{eq:SVD of H2}
\end{align}

\begin{lemma}
The optimal relay beamforming matrix \(\mv W\) for problem $\rm (P1^\prime.1)$ is of the form:
\begin{equation}\label{eq:optimal structure for W}
\mv W=\mv U_1^\dagger\mv B\mv U_2^H+\mv U_1^\dagger\mv C(\mv U_2^{\bot})^H,
\end{equation}
where \(\mv B\in\mathbb{C}^{2\times2}\) and \(\mv C\in\mathbb{C}^{2\times (N_t-2)}\) are two unknown matrices, and \(\mv U_1^{\bot}\), \(\mv U_2^{\bot} \in\mathbb{C}^{N_t\times(N_t-2)}\) satisfy \(\mv U_1^{\bot}(\mv U_1^{\bot})^H=\mv I-\mv U_1\mv U_1^H\), \(\mv U_2^{\bot}(\mv U_2^{\bot})^H=\mv I-\mv U_2\mv U_2^H\), respectively. \label{lemma:optimal structure for W}
\end{lemma}

\begin{IEEEproof}
See Appendix \ref{appendix:proof of lemma:optimal structure for W}.
\end{IEEEproof}

Denote \(\mv U_1^H\tilde{\mv h}_0\), \(\mv U_2^H\mv h_0\), \(\mv U_1^H\mv g_0\) by \(\bar{\tilde{\mv h}}_0\), \(\bar{\mv h}_0\), \(\bar{\mv g}_0\), respectively. We thus simplify \(\vert\widetilde{\mv{h}}^T_0\mv{Wh}_0\vert^2\) and \(\vert\mv g_0^T\mv{Wh}_0\vert^2\) as \(\vert\bar{\tilde{\mv h}}_0^T\mv B\bar{\mv h}_0\vert^2\) and \(\vert\bar{\mv g}_0^T\mv B\bar{\mv h}_0\vert^2\), respectively. Since \(\mv C\) has \(2(N_t-2)\) complex variables, we devise a suboptimal design for \(\mv C\) to reduce the size of variables by \((N_t-2)\). Specifically, let \(\mv C=\mv u^{\prime\bot}\mv v^T\), where \(\mv u^\prime=\bar{\tilde{\mv h}}_0^\dag/\|\bar{\tilde{\mv h}}_0\|\) such that  \(\mv u^{\prime\bot}\mv u^{\prime\bot H}=\mv I-\mv u^\prime\mv u^{\prime H}\). Hence, \(\widetilde{\mv{h}}^T_0\mv{WW}^H\widetilde{\mv{h}}_0^\dagger\), \({\mv{g}}^T_0\mv{WW}^H\mv g_0^\dagger\) and \eqref{eq:transmit power at the relay} can be reduced to \(\|\mv B^H\bar{\tilde{\mv h}}_0^\dagger\|^2\), \(\|\mv B^H\bar{\mv g}_0^\dagger\|^2+\vert\mv v^\dagger\mv u^{\prime\bot H}\bar{\mv g}_0^\dagger\vert^2\) and \(P_s\|\mv B\bar{\mv h}_0\|^2+\sigma_r^2\mathrm{tr}(\mv B^H\mv B)+\sigma_r^2\|\mv v\|^2\), respectively. Then define  \(\mv b={\rm vec}(\mv B)\), \(\bar{\mv f}_1={\rm vec}(\bar{\tilde{\mv h}}_0\bar{\mv h}_0^T)\), \(\bar{\mv f}_2={\rm vec}(\bar{\mv g}_0\bar{\mv h}_0^T)\), \(\mv Y_1^\prime=\bar{\tilde{\mv h}}_0^T\otimes\mv I\), \(\mv Y_2^\prime=\bar{\mv g}_0^T\otimes\mv I\), and \(\mv\Phi^\prime=(\mv I\otimes\mv\Theta^{\prime T})^{1/2}\) with \(\mv\Theta^{\prime}=P_s\bar{\mv h}_0\bar{\mv h}_0^H+\sigma_r^2\mv I\); \(\mv Z=\mv b\mv b^H\), \(\mv V=\mv v\mv v^H\), \(\overline{\mv F}_1=\bar{\mv f}_1^\dagger\bar{\mv f}_1^T\), \(\overline{\mv F}_2=\bar{\mv f}_2^\dagger\bar{\mv f}_2^T\), \(\overline{\mv Y}_1^\prime=\mv Y_1^{\prime H}\mv Y_1^\prime\), \(\overline{\mv Y}_2^\prime=\mv Y_2^{\prime H}\mv Y_2^\prime\), and \(\overline{\mv \Phi}^\prime=\mv \Phi^{\prime H}\mv\Phi^\prime\). The suboptimal design for problem $\rm (P1^\prime.1)$ by ignoring the rank constraints on \(\mv Z\) and \(\mv V\) is thus given by
\begin{subequations}
\begin{align}
 &\!\!\!\! \mathrm{(P1^\prime.1\text{-}sub1\text{-}SDR)}:~ \mathop{\mathtt{max}}_{\boldsymbol{Z},\boldsymbol{V},\{\boldsymbol{Q}_k\},\tau}~~~ P_s{\rm tr}(\overline{\mv F}_1\mv Z)\notag\\
&\!\!\!\! \mathtt{s.t.}~~\sigma^2_r{\rm tr}(\overline {\mv Y}^\prime_1\mv Z)+\sum_{k=1}^K\tilde{\mv h}_k^T\mv Q_k\tilde{\mv h}_k^\dagger+\tau\sigma_b^2=1,\\
&\!\!\!\! P_s{\rm tr}(\overline{\mv F}_2\mv Z)\le\bar\gamma_e\nonumber\\
& \!\!\!\!\left(\sigma^2_r\left({\rm tr}(\overline {\mv Y}^\prime_2\mv Z)+\vert\bar{\mv g}_0^T\mv u^{\prime\bot}\vert^2\mathrm{tr}(\mv V)\right)+\sum_{k=1}^K\mv g_k^T\mv Q_k\mv g_k^\dagger+\tau\sigma_e^2\right),\\
&~~{\rm tr}(\overline{\mv \Phi}^\prime\mv Z)+\sigma_r^2{\rm tr}(\mv Z)+\sigma_r^2\mathrm{tr}(\mv V)\leq \tau P_r,\\
&~~\mathrm{tr}\left(\mv{Q}_k \right)\leq \tau\eta P_s\left\|\mv{h}_k\right\|^2, \; \forall k,\\
&~~\tau\ge 0, \, \mv Q_k\succeq \mv 0, \; \forall k, \,\mv Z\succeq \mv 0, \, \mv V\succeq \mv 0.
\end{align}
\end{subequations}

\begin{remark}
The variables in $\rm (P1^\prime.1\text{-}sub1\text{-}SDR)$, i.e., \(\mv Z\in\mathbb{C}^{4\times4}\), \(\mv V\in\mathbb{C}^{(N_t-2)\times(N_t-2)}\),  are of much reduced size. Further, the reconstruction of \(\mv v^\ast\) from \(\mv V\) can be briefly explained as follows. Given the Lagrangian of $\rm (P1^\prime.1\text{-}sub1\text{-}SDR)$, the KKT conditions with respect to (w.r.t.)~\(\mv V^\ast\) are given by
\begin{align}
(\alpha^\ast\bar\gamma_e\vert\bar{\mv g}_0^T\mv u^{\prime\bot}\vert^2-\beta_0^\ast\sigma_r^2)\mv I +\mv U^\ast &=0, \label{eq:zero derivitive wrt V} \\
\mv U^\ast\mv V^\ast &=0.\label{eq:complementary slackness wrt V}
\end{align}
\end{remark}
Post-multiplying \eqref{eq:zero derivitive wrt V} with \(\mv V^\ast\), we have \((\alpha^\ast\bar\gamma_e\vert\bar{\mv g}_0^T\mv u^{\prime\bot}\vert^2-\beta_0^\ast\sigma_r^2)\mv V^\ast=0\). As a result, if \(\tfrac{\alpha^\ast}{\beta_0^\ast}\neq\tfrac{\sigma_r^2}{\bar\gamma_e\vert\bar{\mv g}_0^T\mv u^{\prime\bot}\vert^2}\), \(\mv V^\ast=\mv 0\); otherwise \(\mv V^\ast=\mv v^\ast\mv v^{\ast H}\), with \(\mv v^\ast=\sqrt{{\rm tr}(\mv V^\ast)}\mv v_0\), where \(\mv v_0\in\mathbb{C}^{(N_t-2)\times1}\) is an arbitrary vector with unit norm. With \(\mv V\) solved, $\rm (P1^\prime.1\text{-}sub1\text{-}SDR)$ reduces to a problem with similar structure as $\rm (P1^\prime.1\text{-}RW\text{-}SDR)$, and the proof for existence of a rank-one \(\mv Z\) can be referred to Proposition \ref{prop:structure of optimal X and its rank-one reconstruction}.

\subsubsection{\textcolor{black}{Zero-forcing}}\label{subsubsec:ZF}
\textcolor{black}{We propose a low-complexity ZF scheme for \(\mathrm{(P1^\prime.1)}\), in which the jamming signal places a null at Bob, and then a semi-closed form solution for \(\mv W\) is derived. In line with the principle of ZF jamming \cite{Zheng2011CJ}, the jamming signal \(\mv n_k\) is designed as \(\mv n_k=\tilde{\mv V}_k\tilde{\mv n}_k\) such that \(\mv I-\tfrac{\tilde{\mv h}_k^\dag\tilde{\mv h}_k^T}{\|\tilde{\mv h}_k\|^2}=\tilde{\mv V}_k\tilde{\mv V}_k^H\), and \(\tilde{\mv n}_k\in\mathbb{C}^{(N_t-1)\times 1}\) is an arbitrary random vector, \(\tilde{\mv n}_k\sim\mathcal{CN}(\mv 0, \tilde{\mv Q}_k)\), \(k=1,\ldots,K\). Thus, given any \(\mv W\), \(\tilde{\mv n}_k\)'s can be optimized to maximize the effect of jamming at Eve by \(\max\limits_{\tilde{\boldsymbol Q}_k}\!\sum_{k=1}^K\mv g_k^T\tilde{\mv V}_k\tilde{\mv Q}_k\tilde{\mv V}_k^H\mv g_k^\dag\), which gives \(\tilde{\mv Q}_k^\ast=\zeta_k^2\tilde{\mv g}_k^\dag\tilde{\mv g}_k^T\), where \(\tilde{\mv g}_k=\tilde{\mv V}_k^T\mv g_k\), and \(\zeta_k=\sqrt{\eta P_s}\|\mv h_k\|/\|\tilde{\mv g}_k\|\) is determined  by \eqref{eq:constraint on amount of AN for each HJ helper}, \(\forall k\). As such, \(\sum_{k=1}^K\mv g_k^T\tilde{\mv V}_k\tilde{\mv Q}_k^\ast\tilde{\mv V}_k^H\mv g_k^\dag\) turns out to be \(\sum_{k=1}^K\eta P_s\|\mv h_k\|^2\|\tilde{\mv g}_k\|^2\), which is denoted by \(q\).}

\textcolor{black}{With fixed \(q\), \(\mathrm{(P1^\prime.1\text{-}RW\text{-}SDR)}\) can be recast as
\begin{subequations}
  \begin{align}
 &\mathrm{(P1^\prime.1\text{-}sub2\text{-}SDR)}:~\mathop{\mathtt{max}}_{\boldsymbol {X},\tau}
~P_s{\rm tr}(\mv F_1\mv X)\nonumber\\
&\mathtt{s.t.}~\sigma^2_r{\rm tr}(\overline {\mv Y}_1\mv X)+\tau\sigma_b^2=1, \label{eq:C1 of ZF}\\
&P_s{\rm tr}(\mv F_2\mv X)\le\bar\gamma_e\left(\sigma^2_r{\rm tr}(\overline {\mv Y}_2\mv X)+\tau q+\tau\sigma_e^2\right), \label{eq:C2 of ZF}\\
&{\rm tr}(\overline{\mv \Phi}\mv X)\leq \tau P_r, \label{eq:C3 of ZF}\\
&\mv X\succeq \mv 0, \, \tau\ge 0.
\end{align}
\end{subequations}}
\textcolor{black}{\begin{proposition}
\(\mathrm{(P1^\prime.1\text{-}sub2\text{-}SDR)}\) must yield a rank-one solution, i.e., \(\mv X^\ast=\mv w\mv w^\ast\), such that \(\mv w^\ast=\mu\mv\nu_{\max}(\mv Z^\ast)\), and
\begin{align}
\mv Z^\ast=&P_s\mv F_1-\lambda^\ast\sigma_r^2\overline{\mv Y}_1-\alpha^\ast P_s\mv F_2\nonumber\\
&+\alpha^\ast\bar\gamma_e\sigma_r^2\overline{\mv Y}_2-\beta_0^\ast\overline{\mv \Phi}, \label{eq:Z for sub2}
\end{align}
where \(\mv \nu_{\max}(\cdot)\) represents the eigenvector corresponding to the largest eigenvalue of the associated matrix, and \(\mu=\sqrt{\tfrac{P_r}{{\rm tr}(\overline{\mv\Phi})\mv\nu_{\max}(\mv Z^\ast)\mv\nu_{\max}^H(\mv Z^\ast)}}\). Also, \(\lambda^\ast\), \(\alpha^\ast\) and \(\beta_0^\ast\) are the optimal dual variables associated with \eqref{eq:C1 of ZF}--\eqref{eq:C3 of ZF}, respectively.\label{prop:semi-closed form}
\end{proposition}}

\begin{IEEEproof}
See Appendix \ref{appendix:proof of prop:semi-closed form}.
\end{IEEEproof}

\textcolor{black}{The only problem in Proposition \ref{prop:semi-closed form} is the dual problem of \(\mathrm{(P1^\prime.1\text{-}sub2\text{-}SDR)}\), which admits a much simpler structure to solve than the primal one.}

\section{Joint AN-AF Beamforming with Imperfect CSI}\label{sec:A Joint Optimization Based on Imperfect CSI}
\subsection{Problem Formulation for Imperfect CSI}\label{subsec:Problem Formulation for Imperfect CSI}
We use a deterministic spherical model \cite{li2011optimal,swindlehurst2012robust} to characterize the resulting CSIs' uncertainties such that
\begin{subequations}\label{eq:uncertainty regions}
\begin{align}
\mathcal{G}_0= &\{\mv g_0\vert\mv g_0=\hat{\mv {g}}_0+\Delta\mv g_0, \Delta\mv g_0^H\mv W_0\Delta\mv g_0\le 1\},\\
\mathcal{G}_k= & \{\mv g_k\vert\mv g_k=\hat{\mv {g}}_k+\Delta\mv g_k, \Delta\mv g_k^H\mv W_k\Delta\mv g_k\le 1\}, \forall k, \\
\tilde{\mathcal H}_0=&\{\tilde{\mv h}_0\vert\tilde{\mv h}_0=\hat{\tilde{\mv h}}_0+\Delta\tilde{\mv h}_0,\Delta\tilde{\mv h}_0^H{\mv W}_0^\prime\Delta\tilde{\mv h}_0\le 1\},\\
\tilde{\mathcal H}_k= &\{\tilde{\mv h}_k\vert\tilde{\mv h}_k=\hat{\tilde{\mv h}}_k+\Delta\tilde{\mv h}_k,\Delta\tilde{\mv h}_k^H\mv W_k^{\prime\prime}\Delta\tilde{\mv h}_k\le 1\}, \forall k, \\
\mathcal{H}_k= & \{\mv h_k\vert\mv h_k=\hat{\mv {h}}_k+\Delta\mv h_k, \Delta\mv h_k^H\mv W_k^{\prime}\Delta\mv h_k\le 1\}, \forall k,
\end{align}
\end{subequations}
where \(\hat{\mv g}_0\), \(\hat{\mv g}_k\)'s, \(\hat{\tilde{\mv h}}_0\),  \(\hat{\tilde{\mv h}}_k\)'s and \(\hat{\mv h}_k\)'s are the estimates of the corresponding channels; \(\Delta\mv g_0\), \(\Delta\mv g_k\)'s, \(\Delta\tilde{\mv h}_0\), \(\Delta\tilde{\mv h}_k\)'s and \(\Delta\mv h_k\)'s are their respective channel errors; the matrices \(\mv W_0\), \(\mv W_k\)'s, \(\mv W_0^\prime\), \(\mv W_k^{\prime\prime}\)'s and \(\mv W_k^\prime\)'s determine the shape of each error region. Without loss of generality (w.l.o.g.), we set \(\mv W_0=\mv I/\epsilon_0\), \(\mv W_0^\prime=\mv I/\epsilon_0^\prime\), \(\mv W_k=\mv I/\epsilon_k\), \(\mv W_k^\prime=\mv I/\epsilon_k^\prime\) and \(\mv W_k^{\prime\prime}=\mv I/\epsilon_k^{\prime\prime}\) for simplicity, where \(\epsilon_0\), \(\epsilon_0^\prime\), \(\epsilon_k\), \(\epsilon_k^\prime\), and \(\epsilon_k^{\prime\prime}\) represent the respective size of the bounded error regions, \(k=1,\ldots, K\).

Accordingly, we denote the robust counterpart for \(\mathrm{(P1^\prime)}\) as
\begin{subequations}
  \begin{align}
 \mathrm{(P2^\prime)}:&~~~\mathop{\mathtt{max}}_{\{\boldsymbol{Q}_k\},\boldsymbol{W}}\mathop{\mathtt{min}}_{{\tilde{\boldsymbol h}_0\in\tilde{\mathcal{H}}_0,\tilde{\boldsymbol{h}}_k\in\tilde{\mathcal H}_k,\forall k}\atop{\boldsymbol g_0\in\mathcal{\boldsymbol G}_0},\tilde{\boldsymbol g}_k\in\tilde{\mathcal G}_k,\forall k}\!\!\!\!\!\!\bar F(\{\mv Q_k\},\mv W)\notag\\
\mathtt{s.t.}& ~~~{\rm tr}\left(\mv W\left(P_s\mv{h}_0\mv{h}^H_0  +\sigma^2_r\mv{I}\right)\mv W^H \right )\leq P_r,\\
& ~~~{\rm tr}\left(\mv{Q}_k \right )\le \eta P_s\!\min\limits_{\boldsymbol h_k\in\mathcal{H}_k,\forall k}\!\left\|\mv{h}_k\right\|^2, \; \forall k,\\
&~~~\mv{Q}_k\succeq \mv 0, \; \forall k.
  \end{align}
\end{subequations}
An equivalent robust reformulation of \(\mathrm{(P1^\prime.2)}\) is given by
\begin{align}
\mathrm{(P2^\prime.2)}: \mathop{\mathtt{max}}_{\bar\gamma_e>0}
 &~~~\frac{1+\hat H(\bar\gamma_e)}{1+\hat F(\bar\gamma_e)}, \label{eq:P(2'.2)}
\end{align}
where \(\hat F(\bar\gamma_e)=\gamma_e\) and  \(\hat H(\bar\gamma_e)\) denotes the optimal value of problem \(\mathrm{(P2^\prime.1)}\) that is given by
\begin{subequations}
\begin{align}
&\mathrm{(P2^\prime.1)}:\notag\\
&\!\!\mathop{\mathtt{max}}_{\boldsymbol{X},\{\boldsymbol{Q}_k\}}\!\mathop{\min}_{{\tilde{\boldsymbol{h}}_k\in\tilde{\mathcal H}_k,\forall k}\atop{\tilde{\boldsymbol h}_0\in\tilde{\mathcal{H}}_0}}
~\frac{P_s{\rm tr}(\mv F_1\mv X)}{\sigma^2_r{\rm tr}(\overline {\mv Y}_1\mv X)+\sum_{k=1}^K\tilde{\mv h}_k^T\mv Q_k\tilde{\mv h}_k^\dagger+\sigma_b^2}\notag\\
&\mathtt{s.t.}\max\limits_{\boldsymbol g_k\in\mathcal{G}_k,\forall k\atop\boldsymbol g_0\in\mathcal{G}_0}\frac{P_s{\rm tr}(\mv F_2\mv X)}{\sigma^2_r{\rm tr}(\overline {\mv Y}_2\mv X)+\sum_{k=1}^K\mv g_k^T\mv Q_k\mv g_k^\dagger+\sigma_e^2}\le\bar\gamma_e, \label{eq:epigraph reformulation on gamma_e}\\
&{\rm tr}(\overline{\mv \Phi}\mv X)\leq P_r, \label{eq:constraint on the relay power}\\
&{\rm tr}\left(\mv{Q}_k \right )\le \eta P_s\!\min\limits_{\boldsymbol h_k\in\mathcal{H}_k,\forall k}\!\left\|\mv{h}_k\right\|^2, \; \forall k, \label{eq:constraints on the jamming power}\\
& {\rm rank}(\mv X)=1, \label{eq:constraint on the rank}\\
& \mv X\succeq \mv 0, \, \mv Q_k\succeq \mv 0, \; \forall k. \label{eq:PSD constriants}
\end{align}
\end{subequations}
As stated in Lemma \ref{lemma:P1'.2 same optimal value}, similarly, \(\mathrm{(P2^\prime)}\) can be proved to have the same optimal value as \(\mathrm{(P2^\prime.2)}\) and the same optimal solution as \(\mathrm{(P2^\prime.1)}\) when \(\bar\gamma_e\) takes its optimal value. As a result, \(\mathrm{(P2^\prime)}\) can be solved in a two-stage fashion as well. Specifically, given any \(\bar\gamma_e\), we first solve \(\mathrm{(P2^\prime.1)}\) to obtain \(\hat H(\bar\gamma_e)\) and then search for the optimal \(\bar\gamma_e\) to \(\mathrm{(P2^\prime.2)}\).


\subsection{Solutions to $\rm (P2^\prime.1)$}
By ignoring \eqref{eq:constraint on the rank}, \(\mathrm{(P2^\prime.1)}\) is recast as
\begin{subequations}
  \begin{align}
    &\mathrm{(P2^\prime.1\text{-}RW\text{-}SDR\text{-}Eqv)}:\notag\\
&\!\!\mathop{\mathtt{max}}_{\boldsymbol{X},\{\boldsymbol{Q}_k\}}\!\mathop{\min}_{{\tilde{\boldsymbol{h}}_k\in\tilde{\mathcal H}_k,\forall k}\atop{\tilde{\boldsymbol h}_0\in\tilde{\mathcal{H}}_0}}
~\frac{P_s{\rm tr}(\mv F_1\mv X)}{\sigma^2_r{\rm tr}(\overline {\mv Y}_1\mv X)+\sum_{k=1}^K\tilde{\mv h}_k^T\mv Q_k\tilde{\mv h}_k^\dagger+\sigma_b^2}\notag\\
&\mathtt{s.t.}\eqref{eq:epigraph reformulation on gamma_e}-\eqref{eq:constraints on the jamming power}, \, \eqref{eq:PSD constriants}.
  \end{align}\label{eq:(P2'.1-RW-SDR-Eqv)}
\end{subequations}

It is worth noting that due to the rank-one relaxation of \(\mathrm{(P2^\prime.1\text{-}RW\text{-}SDR\text{-}Eqv)}\), solution provided by \(\mathrm{(P2^\prime.1\text{-}RW\text{-}SDR\text{-}Eqv)}\) in general yields an upper-bound for \(\hat H(\bar\gamma_e)\), which may not be achievable. However, in the sequel we insist on solving \(\mathrm{(P2^\prime.1\text{-}RW\text{-}SDR\text{-}Eqv)}\) that is regarded as an upper-bound benchmark for our proposed problem detailed later in this subsection.

\subsubsection{\textcolor{black}{Solutions to $\mathrm{(P2^\prime.1\text{-}RW\text{-}SDR\text{-}Eqv)}$}}
\textcolor{black}{To make the ``max-min'' objective function of \eqref{eq:(P2'.1-RW-SDR-Eqv)} tractable, we first rewrite \eqref{eq:(P2'.1-RW-SDR-Eqv)} by the equivalent epigraph formulation as
\begin{subequations}
  \begin{align}
  &~\mathrm{(P2^\prime.1\text{-}RW\text{-}SDR\text{-}Eqv)}:\nonumber\\
&\!\!\mathop{\mathtt{max}}_{\boldsymbol{X},\{\boldsymbol{Q}_k\},\delta}~\delta\nonumber\\
&~\mathtt{s.t.}\min\limits_{{\tilde{\boldsymbol{h}}_k\in\tilde{\mathcal H}_k,\forall k}\atop{\tilde{\boldsymbol h}_0\in\tilde{\mathcal{H}}_0}}
\frac{P_s{\rm tr}(\mv F_1\mv X)}{\sigma^2_r{\rm tr}(\overline {\mv Y}_1\mv X)+\sum_{k=1}^K\tilde{\mv h}_k^T\mv Q_k\tilde{\mv h}_k^\dagger+\sigma_b^2}\ge\delta, \label{eq:epigraph reformulation on delta}\\
&~\eqref{eq:epigraph reformulation on gamma_e}-\eqref{eq:constraints on the jamming power}, \, \eqref{eq:PSD constriants}.
  \end{align}
\end{subequations}
As there are potentially infinite number of constraints in \eqref{eq:epigraph reformulation on delta}, \eqref{eq:epigraph reformulation on gamma_e}, and \eqref{eq:constraints on the jamming power}, they are semi-indefinite and thus intractable. In the following, we equivalently transform these constraints to tractable ones using {\em S-Procedure} and a generalized {\em S-Procedure} given in Lemmas~\ref{lemma:S-Procedure} and \ref{lemma:LMI from robust block QMI}, respectively.
}
\begin{lemma}\label{lemma:S-Procedure} ({\em S-Procedure} \cite{boyd2004convex}) Let \(f_m(\mv x)\), \(m=1,2\) be defined as
\begin{equation}\label{eq:f_m(x)}
f_m(\mv x)=\mv x^H\mv A_m\mv x+2\Re\{\mv b_m^H\mv x\}+c_m,
\end{equation}
where \(\mv A_m=\mv A_m^H\in\mathbb{C}^{N\times N}\), \(\mv b_m\in\mathbb{C}^{N\times 1}\) and \(c_m\in\mathbb{R}\), and $\Re$ gives the real part of the input entity. Then the implication \(f_1(\mv x)\ge 0\Rightarrow f_2(\mv x)\ge 0\) holds if and only if there exists $\delta\ge 0$ such that
\begin{equation}\label{eq:S-Procedure}
\begin{bmatrix} \mv A_2 & \mv b_2\\
\mv b_2^H & c_2\end{bmatrix}-\delta\begin{bmatrix}\mv A_1 &\mv b_1\\ \mv b_1^H &c_1 \end{bmatrix}\succeq \mv 0,
\end{equation}
provided there exists a point \(\hat{\mv x}\) such that \(f_m(\hat{\mv x})>0\), \(m=1,2\).
\end{lemma}

\begin{lemma}\label{lemma:LMI from robust block QMI}
(\cite[\em Theorem 3.5]{luo2004multivariate}) The robust block quadratic matrix inequality (QMI),
 \begin{align}\label{eq:robust block QMI}
 \begin{bmatrix}
\mv H & \mv F+\mv G\mv X\\
 (\mv F+\mv G\mv X)^H & \mv C+\mv X^H\mv B+\mv B^H\mv X+\mv X^H\mv A\mv X
\end{bmatrix}\succeq\mv 0, \,\nonumber\\
\!\!\!\!\!\!\!\!\!\! \mbox{for all}\,  \mv I-\mv X^H\mv D\mv X\succeq \mv 0,
\end{align}
is equivalent to
\begin{equation}\label{eq:LMI from robust block QMI}
\exists t\ge 0, \, \mbox{such that} \, \begin{bmatrix}
\mv H & \mv F & \mv G\\
\mv F^H & \mv C & \mv B^H\\
\mv G^H & \mv B & \mv A
\end{bmatrix} -t\begin{bmatrix}\mv 0 & \mv 0 & \mv 0 \\\mv 0 & \mv I & \mv 0\\\mv 0 &\mv 0 &-\mv D \end{bmatrix}\succeq\mv 0.
 \end{equation}
\end{lemma}

\textcolor{black}{First, by rearranging terms, \eqref{eq:epigraph reformulation on delta} can be equivalently transformed into the following linear form:
\begin{multline}\label{eq:rearranging on delta}
\min\limits_{{\tilde{\boldsymbol{h}}_k\in\tilde{\mathcal H}_k,\forall k}\atop{\tilde{\boldsymbol h}_0\in\tilde{\mathcal{H}}_0}}\!\!P_s{\rm tr}(\mv F_1\mv X)-\delta\sigma_r^2{\rm tr}(\overline{\mv Y}_1\mv X)\\
-\delta\sum_{k=1}^K\tilde{\mv h}_k^T\mv Q_k\tilde{\mv h}_k^\dag-\delta\sigma_b^2\ge 0.
\end{multline}
Recalling the following matrix equalities in line with our definition of \({\rm vec}(\cdot)\) operation:
\begin{align}
{\rm tr}(\mv A\mv B^T)&={\rm vec}^T(\mv A){\rm vec}(\mv B), \\
{\rm vec}(\mv A\mv X\mv B)&=(\mv A\otimes\mv B^T){\rm vec}(\mv X), \\
 (\mv A\otimes\mv B)^T&=\mv A^T\otimes\mv B^T,
\end{align}
it follows that
\begin{align}
{\rm tr}(\mv F_1\mv X)&=\tilde{\mv h}^T(\mv h_0\otimes\mv I)X(\mv h_0^H\otimes\mv I)\tilde{\mv h}^\dag,\\
{\rm tr}(\overline{\mv Y}_1\mv X)&=\tilde{\mv h}^T(\mv I\otimes\mv X)\tilde{\mv h}^\dag, \label{eq:maniuplation on Y_1}
\end{align}
where \(\tilde{\mv h}\in\mathbb{C}^{N_t^3\times1}={\rm vec}(\tilde{\mv h}_0^T\otimes\mv I)\). The equivalent channel model for \(\tilde{\mv h}\) is given by \(\tilde{\mv h}=\hat{\tilde{\mv h}}+\Delta\tilde{\mv h}\), where \(\|\Delta\tilde{\mv h}\|^2\le N_t\epsilon_0^\prime\) (c.f.~\eqref{eq:uncertainty regions}). By introducing \(\mv X^{\prime\prime}=(\mv h_0\otimes\mv I)\mv X(\mv h_0^H\otimes\mv I)\) and  \(\mv X^\prime=\mv I\otimes\mv X\), \eqref{eq:rearranging on delta} can thus be recast as
\begin{multline}\label{eq:semi-indefinite form of tilde_h for S-procedure}
\!\!\!\!\!\!\!\!\min\limits_{{\tilde{\boldsymbol{h}}_k\in\tilde{\mathcal H}_k,\forall k}\atop{\tilde{\boldsymbol h}_0\in\tilde{\mathcal{H}}_0}}\!\!\Delta\tilde{\mv h}^T(P_s\mv X^{\prime\prime}-\delta\sigma_r^2\mv X^\prime)\Delta\tilde{\mv h}^\dag+2\Re\{\Delta\tilde{\mv h}^T(P_s\mv X^{\prime\prime}-\delta\sigma_r^2\mv X^\prime)\hat{\tilde{\mv h}}^\dag\}\\
-\delta\sum_{k=1}^K\tilde{\mv h}_k^T\mv Q_k\tilde{\mv h}_k^\dag-\delta\sigma_b^2\ge 0.
\end{multline}
Hence, according to Lemma~\ref{lemma:S-Procedure}, the implication \(\|\Delta\tilde{\mv h}\|^2\le N_t\epsilon_0^\prime\Rightarrow\eqref{eq:semi-indefinite form of tilde_h for S-procedure}\) holds if and only if there exists \(w^{(0)}\ge 0\) such that the following LMI holds:
\begin{align}
\begin{bmatrix}\mv H_1 & \mv F_1\\
\mv F_1^H & c_1 \end{bmatrix}\succeq\mv 0, \label{eq:LMI of eqv obj for S-Procedure}
\end{align}
where $\mv H_1=P_s\mv X^{\prime\prime}-\delta\sigma_r^2\mv X^\prime+w^{(0)}\mv I$, $\mv F_1=(P_s\mv X^{\prime\prime}-\delta\sigma_r^2\mv X^\prime)\hat{\tilde{\mv h}}^\dag$ and
$c_1=\hat{\tilde{\mv h}}^T(P_s\mv X^{\prime\prime}-\delta\sigma_r^2\mv X^\prime)\hat{\tilde{\mv h}}^\dag-\delta\sum_{k=1}^K{\tilde{\mv h}}_k^T\mv Q_k{\tilde{\mv h}}_k^\dag-\delta\sigma_b^2-w^{(0)}N_t\epsilon_0^\prime$.
Now, \eqref{eq:epigraph reformulation on delta} has been equivalently reformulated as \eqref{eq:LMI of eqv obj for S-Procedure}. To further cope with channel uncertainties with regards to \(\tilde{\mv h}_k\)'s such that \eqref{eq:LMI of eqv obj for S-Procedure} holds for \(\tilde{\mv h}_k\in\tilde{\mathcal H}_k\), \(k=1,\ldots,K\), we need the following proposition.}
\textcolor{black}{\begin{proposition}
The semi-indefinite constraint of \eqref{eq:semi-indefinite form of tilde_h for S-procedure} can be equivalently recast as the following block matrix inequality:
\begin{align}
\begin{bmatrix} \mv H_1^{(K)} & \mv F_1^{(K)} & \mv G_1^{(K)}\\ \mv F_1^{(K)H} & c_1^{(K)} & \mv B_1^{(K)H}\\ \mv G_1^{(K)H} & \mv B_1^{(K)} & \mv A_1^{(K)}\end{bmatrix}-w^{(K)}\begin{bmatrix}\mv 0 & \mv 0 & \mv 0\\ \mv 0 & 1 & \mv 0\\ \mv 0 & \mv 0 & \frac{-\mv I}{\epsilon_K^{\prime\prime}} \end{bmatrix}\succeq\mv 0,\label{eq:LMI reformulation on delta}
\end{align} where \(\mv H_1^{(K)}\), \(\mv F_1^{(K)}\) and \(c_1^{(K)}\) are recursively given by
\begin{align}
 \mv H_1^{(k)}&=\left\{\begin{array}{ll}
 \begin{bmatrix}\mv A_1^{(k-1)}+\frac{w^{(k-1)}}{\epsilon_{k-1}^{\prime\prime}}\mv I & \mv G_1^{(k-1)H}\\ \mv G_1^{(k-1)} & \mv H_1^{(k-1)}\end{bmatrix}, & k>1;\\
P_s\mv X^{\prime\prime}-\delta\sigma_r^2\mv X^\prime+w^{(0)}\mv I, & k=1,
  \end{array}\right.\\
 \mv F_1^{(k)}&=\left\{\begin{array}{ll}
  \begin{bmatrix}\mv B_1^{(k-1)}\\ \mv F_1^{(k-1)} \end{bmatrix},& k>1;\\
(P_s\mv X^{\prime\prime}-\delta\sigma_r^2\mv X^\prime)\hat{\tilde{\mv h}}^\dag,& k=1,
  \end{array}\right.
\end{align}
\(c_1^{(k)}=\hat{\tilde{\mv h}}^T(P_s\mv X^{\prime\prime}-\delta\sigma_r^2\mv X^\prime)\hat{\tilde{\mv h}}^\dag-\delta\sum_{j=1}^k\hat{\tilde{\mv h}}_j^T\mv Q_j\hat{\tilde{\mv h}}_j^\dag-\delta\sum_{i=k+1}^K{\tilde{\mv h}}_i^T\mv Q_i{\tilde{\mv h}}_i^\dag-\delta\sigma_b^2-w^{(0)}N_t\epsilon_0^\prime-\sum_{l=1}^{k-1}w^{(l)}\), \(k=1,\ldots,K\). In addition, \(\mv G_1^{(k)}\in\mathbb{C}^{(N_t^3+(k-1)N_t)\times N_t}=\mv 0\), \(\mv B_1^{(k)}=-\delta\mv Q_k\hat{\tilde{\mv h}}_k^\dag\), \(\mv A_1^{(k)}=-\delta\mv Q_k\), \(k=1,\ldots,K\), and \(\{w^{(k)}\ge 0\}\) denote pertinent auxiliary variables. \label{prop:eqv LMI wrt tilde_h1 till tilde_hK}
\end{proposition}}

\textcolor{black}{\begin{IEEEproof}
See Appendix \ref{appendix:proof of prop:eqv LMI wrt tilde_h1 till tilde_hk}.
\end{IEEEproof}}

\textcolor{black}{Next, \eqref{eq:epigraph reformulation on gamma_e} is rewritten as
\begin{multline}\label{eq:linear epigraph reformulation on gamma_e}
\max\limits_{\boldsymbol g_k\in\mathcal{G}_k,\forall k\atop\boldsymbol g_0\in\mathcal{G}_0}\!\!\mv g^T\left(P_s\mv X^{\prime\prime}-\bar\gamma_e\sigma_r^2\mv X^\prime\right)\mv g^\dag-\bar\gamma_e\sum_{k=1}^K\mv g_k^T\mv Q_k\mv g_k^\dag\\
-\bar\gamma_e\sigma_e^2\le 0,
\end{multline}
where \(\mv g\in\mathbb{C}^{N_t^2\times 1}={\rm vec}(\mv g_0^T\otimes\mv I)\) and the equivalent imperfect channel model is given by \(\mv g=\hat{\mv g}+\Delta\mv g\) such that \(\|\Delta\mv g\|^2\le N_t\epsilon_0\).
}
\textcolor{black}{\begin{proposition}
The semi-indefinite constraint of \eqref{eq:linear epigraph reformulation on gamma_e} is satisfied if and only if there exists \(v^{(k)}\ge 0\), \(k=1,\ldots,K\), such that the following block matrix inequality holds:
\begin{align}
\begin{bmatrix} \mv H_2^{(K)} & \mv F_2^{(K)} & \mv G_2^{(K)}\\ \mv F_2^{(K)H} & c_2^{(K)} & \mv B_2^{(K)H}\\ \mv G_2^{(K)H} & \mv B_2^{(K)} & \mv A_2^{(K)}\end{bmatrix}-v^{(K)}\begin{bmatrix}\mv 0 & \mv 0 & \mv 0\\ \mv 0 & 1 & \mv 0\\ \mv 0 & \mv 0 & \frac{-\mv I}{\epsilon_K} \end{bmatrix}\succeq\mv 0, \label{eq:LMI reformulation on gamma_e}
\end{align}where \(\mv H_2^{(K)}\), \(\mv F_2^{(K)}\) and \(c_2^{(K)}\) are recursively given by
\begin{align}
 \mv H_2^{(k)}&=\left\{\begin{array}{ll}
 \begin{bmatrix}\mv A_2^{(k-1)}+\frac{v^{(k-1)}}{\epsilon_{k-1}}\mv I & \mv G_2^{(k-1)H}\\ \mv G_2^{(k-1)} & \mv H_2^{(k-1)}\end{bmatrix}, & k>1;\\
-P_s\mv X^{\prime\prime}+\bar\gamma_e\sigma_r^2\mv X^\prime+v^{(0)}\mv I, & k=1,
  \end{array}\right.\\
 \mv F_2^{(k)}&=\left\{\begin{array}{ll}
  \begin{bmatrix}\mv B_2^{(k-1)}\\ \mv F_2^{(k-1)} \end{bmatrix},& k>1;\\
(-P_s\mv X^{\prime\prime}+\bar\gamma_e\sigma_r^2\mv X^\prime)\hat{\mv g}^\dag,& k=1,
  \end{array}\right.
\end{align}
\begin{multline}
c_2^{(k)}=\hat{{\mv g}}^T(-P_s\mv X^{\prime\prime}+\bar\gamma_e\sigma_r^2\mv X^\prime)\hat{{\mv g}}^\dag+\bar\gamma_e\sum_{j=1}^k\hat{{\mv g}}_j^T\mv Q_j\hat{{\mv g}}_j^\dag+\\
\bar\gamma_e\sum_{i=k+1}^K{{\mv g}}_i^T\mv Q_i{{\mv g}}_i^\dag+\bar\gamma_e\sigma_e^2-v^{(0)}N_t\epsilon_0-\sum_{l=1}^{k-1}v^{(l)},
\end{multline}
\(k=1,\ldots,K\). Also, \(\mv G_2^{(k)}=\mv G_1^{(k)}\), \(\mv B_2^{(k)}=\bar\gamma_e\mv Q_k\hat{\mv g}_k^\dag\), and \(\mv A_2^{(k)}=\bar\gamma_e\mv Q_k\), \(k=1,\ldots,K\). \label{prop:eqv LMI wrt g1 till gK}
\end{proposition}}

\textcolor{black}{\begin{IEEEproof}
See Appendix \ref{appendix:proof of prop:eqv LMI wrt g1 till gk}.
\end{IEEEproof}}

Last, we rewrite \eqref{eq:constraints on the jamming power} to facilitate the robust optimization against the errors introduced by \(\Delta\mv h_k\)'s. By applying Lemma \ref{lemma:S-Procedure}, \eqref{eq:constraints on the jamming power} holds if and only if there exists \(\mu_k\ge 0\), \(k=1,\ldots,K\), such that the following LMI constraint is met:
\begin{align}\label{eq:eqv LMI of available AN for S-Procedure}
\begin{bmatrix}\eta P_s\mv I+\mu_k\mv I & \eta P_s\hat{\mv h}_k\\
\eta P_s\hat{\mv h}_k^H & \eta P_s\|\hat{\mv h}_k\|_2^2-{\rm tr}(\mv Q_k)-\mu_k\epsilon_k^\prime \end{bmatrix}\succeq\mv 0, \; \forall k.
\end{align}

As such, \(\mathrm{(P2^\prime.1\text{-}RW\text{-}SDR\text{-}Eqv)}\) is now simplified as
\begin{align}
\mathrm{(P2^\prime.1\text{-}RW\text{-}SDR\text{-}Eqv)}:&~~~\mathop{\mathtt{max}}_{\boldsymbol{X},\{\boldsymbol{Q}_k\},\delta}\delta\nonumber\\
\mathtt{s.t.}&~~~\eqref{eq:LMI reformulation on delta}, \eqref{eq:LMI reformulation on gamma_e}, \eqref{eq:eqv LMI of available AN for S-Procedure},
\eqref{eq:constraint on the relay power}, \eqref{eq:PSD constriants}.\nonumber
\end{align}

\textcolor{black}{Because of the non-convex term such as \(\delta\mv X^\prime\) in \eqref{eq:LMI reformulation on delta}, problem \(\mathrm{(P2^\prime.1\text{-}RW\text{-}SDR\text{-}Eqv)}\) remains very hard to solve. We thus use the bisection method \cite{boyd2004convex} w.r.t. \(\delta\) to solve it. However, using bisection in addition to solving \(\mathrm{(P2^\prime.2)}\) by one-dimension search over \(\bar\gamma_e\) may lead to very high complexity. As a result, we propose an alternative problem to approximate \(\hat H(\bar\gamma_e)\).}

\subsubsection{Solutions to $\mathrm{(P2^\prime.1\text{-}RW\text{-}SDR)}$}
We propose to approximate \(\hat H(\bar\gamma_e)\) by the optimum value of the following problem.
\begin{subequations}
\begin{align}
&\mathrm{(P2^\prime.1\text{-}RW\text{-}SDR)}:~
\mathop{\mathtt{max}}_{\boldsymbol{X},\{\boldsymbol{Q}_k\},\tau} \min\limits_{\tilde{\boldsymbol{h}}_0\in\tilde{\mathcal H}_0}P_s{\rm tr}(\boldsymbol{F}_1\boldsymbol{X})\label{eq:robust obj for C-O transfomrmation}\\
& \mathtt{s.t.} \max\limits_{{\tilde{\boldsymbol{h}}_k\in\tilde{\mathcal H}_k,\forall k}\atop {\tilde{\boldsymbol{h}}_0\in\tilde{\mathcal H}_0}}\sigma^2_r{\rm tr}(\overline {\mv Y}_1\mv X)\!+\!\sum_{k=1}^K\tilde{\mv h}_k^T\mv Q_k\tilde{\mv h}_k^\dagger\!+\!\tau\sigma_b^2 \!\le\! 1, \label{eq:robust constraint on C-O transformation}\\
&\max\limits_{{\boldsymbol{g}_k\in\mathcal{G}_k,\forall k}\atop {\boldsymbol{g}_0\in\mathcal{G}_0}}\frac{P_s{\rm tr}(\mv F_2\mv X)}{\sigma^2_r{\rm tr}(\overline {\mv Y}_2\mv X)+\sum_{k=1}^K\mv g_k^T\mv Q_k\mv g_k^\dagger+\tau\sigma_e^2}\le\bar\gamma_e, \label{eq:robust constraint on SINR of Eve}\\
&{\rm tr}(\overline{\mv \Phi}\mv X)\leq \tau P_r, \label{eq:robust constraint on transmit power of the relay}\\
&{\rm tr}\left(\mv{Q}_k \right )\le \tau\eta P_s\!\min\limits_{\mv h_k\in\mathcal{H}_k,\forall k}\!\left\|\mv{h}_k\right\|^2, \; \forall k, \label{eq:robust constraint on amount of AN for each HJ helper}\\
&\mv X\succeq \mv 0, \, \mv Q_k\succeq \mv 0, \; \forall k, \, \tau\ge 0.\label{eq:robust constraint on PSD}
\end{align}
\end{subequations}

\textcolor{black}{\begin{remark}
  It is worth noting that as the numerator and the denominator of the objective function in \(\mathrm{(P2^\prime.1)}\) are coupled by common uncertainty \(\tilde{\mv h}_0\), Charnes-Cooper transformation, in general, cannot be applied to realize equivalent decoupling. 
  As a result, \(\mathrm{(P2^\prime.1\text{-}RW\text{-}SDR)}\) yields a more conservative approximation for \(\hat H(\bar\gamma_e)\) than \(\mathrm{(P2^\prime.1\text{-}RW\text{-}SDR\text{-}Eqv)}\). However, considering that   \(\mathrm{(P2^\prime.1\text{-}RW\text{-}SDR)}\) needs to be solved only once for given \(\bar\gamma_e\) in contrast with \(\mathrm{(P2^\prime.1\text{-}RW\text{-}SDR\text{-}Eqv)}\) requring isection over \(\delta\), we exploit it in the sequel. The effectiveness of this approximation will be evaluated in Section~\ref{subsec:The Imperfect CSI Case}.
\end{remark}}

To proceed, we rewrite $\rm (P2^\prime.1\text{-}RW\text{-}SDR)$ as
\begin{subequations}
\begin{align}
&\mathrm{(P2^\prime.1\text{-}RW\text{-}SDR)}: \mathop{\mathtt{max}}_{\boldsymbol{X},\{\boldsymbol{Q}_k\},\delta,\tau}&~\delta~\notag\\
\mathtt{s.t.}&~\min\limits_{\tilde{\boldsymbol{h}}_0\in\tilde{\mathcal H}_0}P_s{\rm tr}(\mv F_1\mv X)\ge\delta,\label{eq:robust constraint on delta}\\
&~\eqref{eq:robust constraint on C-O transformation}\text{--}\eqref{eq:robust constraint on PSD}.
\end{align}
\end{subequations}

First, by rewriting \(\mv F=\mv f_1^\dag\mv f_1^T\), where \(\mv f_1=\hat{\mv f}_1+\Delta\mv f_1\), in line with Lemma \ref{lemma:S-Procedure}, the implication \(\|\Delta{\mv f}_1\|^2\le\|\mv h_0\|^2\epsilon_0^\prime\Rightarrow\eqref{eq:robust constraint on delta}\) holds if and only if there exists \(s^{\prime(0)}\ge 0\) such that the following LMI constraint is satisfied:
\begin{equation}\label{eq:LMI of obj for S-Procedure}
\begin{bmatrix}P_s\mv X+s^{\prime(0)}\mv I & P_s\mv X\hat{\mv f}_1^\dag\\
P_s\hat{\mv f}_1^T\mv X & P_s\hat{\mv f}_1^T\mv X\hat{\mv f}_1^\dag-s^{\prime(0)}\epsilon_0^\prime\|\mv h_0\|_2^2-\delta \end{bmatrix}\succeq\mv 0.
\end{equation}

Next, as \({\rm tr}(\overline{\mv Y}_1\mv X)=\mv y_1^T\mv X^\prime\mv y_1^\dag\) (c.f.~\eqref{eq:maniuplation on Y_1}), where \(\mv y_1={\rm vec}(\tilde{\mv h}_0^T\otimes\mv I)\), after some manipulation,  \eqref{eq:robust constraint on C-O transformation} holds if and only if there exists \(s^{\prime\prime(0)}\ge 0\) such that
\begin{equation}\label{eq:LMI of y1 for S-Procedure}
 \begin{bmatrix} s^{\prime\prime(0)}\mv I-\sigma_r^2\mv X^{\prime}  \! & \!  -\sigma_r^2\mv X^{\prime}\hat{\mv y}_1^\dag\\
-\sigma_r^2\hat{\mv y}_1^T\mv X^\prime \! & \! c \end{bmatrix}\succeq\mv 0,
\end{equation}
where $ c = -\sigma_r^2\hat{\mv y}_1^T\mv X^\prime\hat{\mv y}_1^\dag-\sum_{k=1}^K\tilde{\mv h}_k^T\mv Q_k\tilde{\mv h}_k^\dag-\tau\sigma_b^2+1-s^{\prime\prime(0)}N_t\epsilon_0^\prime $.
Then \eqref{eq:robust constraint on C-O transformation} can be rewritten as \begin{align}
\eqref{eq:LMI of y1 for S-Procedure}\ \mbox{for}\ \tilde{\mv h}_k\in\tilde{\mathcal H}_k, \; \forall k, \label{eq:robust reformulation on C-O transformation}
\end{align} which is handled by the following proposition.

\begin{proposition}
The semi-indefinite constraints in \eqref{eq:robust reformulation on C-O transformation} can be replaced by the following LMI constraint:
\begin{equation}\label{eq:LMI wrt tilde_h1 till tilde_hK}
\begin{bmatrix}\bar{\mv H}^{(K)}&\bar{\mv F}^{(K)} &\bar{\mv G}^{(K)} \\
\bar{\mv F}^{(K)H} & \bar c^{(K)}  & \bar{\mv B}^{(K)H} \\
\bar{\mv G}^{(K)H}& \bar{\mv B}^{(K)}& \bar{\mv A}^{(K)}
\end{bmatrix}-s^{\prime\prime(K)}\begin{bmatrix} \mv 0 & \mv 0 &\mv 0\\\mv 0 &1 &\mv 0\\ \mv 0&\mv 0 &\frac{-\mv I}{\epsilon_K^{\prime\prime}} \end{bmatrix}\succeq {\bf 0},
\end{equation}
where \(\bar{\mv H}^{(K)}\) and \(\bar{\mv F}^{(K)}\) are recursively given by
\begin{align}
\!\!\!\begin{cases}\bar{\mv H}^{(k)}\!\!=\!\!\begin{bmatrix}\bar{\mv A}^{(k-1)}\!+\!\frac{s^{\prime\prime(k-1)}\mv I}{\epsilon_{k-1}^{\prime\prime}} \!&\! \bar{\mv G}^{(k-1)H}\\ \bar{\mv G}^{(k-1)}&\bar{\mv H}^{(k-1)} \end{bmatrix}, \bar{\mv F}^{(k)}\!\!=\!\!\begin{bmatrix}\bar{\mv B}^{(k-1)} \\ \bar{\mv F}^{(k-1)}\end{bmatrix}\\
k=2,\ldots,K;\\
\bar{\mv H}^{(1)}=s^{\prime\prime(0)}\mv I-\sigma_r^2\mv X^{\prime}, \bar{\mv F}^{(1)}=-\sigma_r^2\mv X^\prime\hat{\mv y}_1^\dag,
\end{cases}\label{eq:recursive relation of bar Hk and bar Fk}
\end{align}
where
$\bar{\mv G}^{(k)}=\mv G_1^{(k)}$, $\bar{\mv B}^{(k)}=-\mv Q_k\hat{\tilde{\mv h}}_k^\dag$, $\bar{\mv A}^{(k)}=-\mv Q_k$, $\bar c^{(k)}=-\sigma_r^2\hat{\mv y}_1^T\mv X^\prime\hat{\mv y}_1^\dag-\Sigma_{j=1}^k\hat{\tilde{\mv h}}_j^T\mv Q_j\hat{\tilde{\mv h}}_j^\dag-\Sigma_{i=k+1}^K\tilde{\mv h}_i^T\mv Q_i\tilde{\mv h}_i^\dag-\tau\sigma_b^2+1-s^{\prime\prime(0)}N_t\epsilon_0^\prime-\Sigma_{l=1}^{k-1}s^{\prime\prime(l)}$, \(k=1,\ldots,K\), and \(\{s^{\prime\prime(k)}\ge 0\}\) denote the auxiliary variables. \label{prop:LMI wrt tilde_h1 till tilde_hK}
\end{proposition}

\begin{IEEEproof}
See Appendix \ref{appendix:proof of prop:LMI wrt tilde_h1 till tilde_hK}.
\end{IEEEproof}


\begin{proposition}
The constraint in \eqref{eq:robust constraint on SINR of Eve} is guaranteed if and only if there exists \(s^{(k)}\ge 0\), \(k=1,\ldots,K\), such that the following LMI holds:
\begin{equation}\label{eq:LMI wrt g1 till gK}
\begin{bmatrix}
\mv H^{(K)}&\mv F^{(K)} &\mv G^{(K)} \\
\mv F^{(K)H} & c^{\prime(K)}  & \mv B^{\prime(K)H} \\
\mv G^{(K)H}& \mv B^{\prime(K)}& \mv A^{\prime(K)}
\end{bmatrix}-s^{(K)}\begin{bmatrix} \mv 0 & \mv 0 &\mv 0\\\mv 0 &1 &\mv 0\\ \mv 0&\mv 0 &\frac{-\mv I}{\epsilon_K} \end{bmatrix}\succeq {\bf 0},
\end{equation}
where \(\mv H^{(k)}\) and \(\mv F^{(k)}\) are recursively given by
\begin{align}
\!\!\!\begin{cases}\mv H^{(k)}\!\!=\!\!\begin{bmatrix}
\mv A^{\prime(k-1)}\!+\!\frac{s^{(k-1)}\mv I}{\epsilon_{k-1}}\! &\!\mv G^{(k-1)H} \\
\mv G^{(k-1)}& \mv H^{(k-1)}
\end{bmatrix}, \mv F^{(k)}\!\!=\!\!\begin{bmatrix} \mv B^{\prime(k-1)}\\ \mv F^{(k-1)} \end{bmatrix}\\
k=2,\ldots,K;\\
\mv H^{(1)}=-P_s\mv X^{\prime\prime}+\bar\gamma_e\sigma_r^2\mv X^\prime+s^{(0)}\mv I,\\
\mv F^{(1)}=\left(-P_s\mv X^{\prime\prime}+\bar\gamma_e\sigma_r^2\mv X^\prime\right )\hat{\mv g}^\dag,
\end{cases}\label{eq:recursive relation of Hk and Fk}
\end{align}
 in which \(\mv G^{(k)}=\mv G_1^{(k)}\), \(\mv B^{\prime(k)}=\bar\gamma_e\mv Q_k\hat{\mv g}_k^\dagger\), \(\mv A^{\prime(k)}=\bar\gamma_e\mv Q_k\), $c^{\prime(k)}=\hat{{\mv g}}^T(-P_s\mv X^{\prime\prime}+\bar\gamma_e\sigma_r^2\mv X^\prime)\hat{{\mv g}}^\dag+\bar\gamma_e\sum_{j=1}^k\hat{{\mv g}}_j^T\mv Q_j\hat{{\mv g}}_j^\dag+\bar\gamma_e\sum_{i=k+1}^K{{\mv g}}_i^T\mv Q_i{{\mv g}}_i^\dag+
\bar\gamma_e\tau\sigma_e^2-s^{(0)}N_t\epsilon_0-\sum_{l=1}^{k-1}s^{(l)}$, $k=1, \ldots, K$,
and \(\{s^{(k)}\ge 0\}\) denote the auxiliary variables. \label{prop:LMI wrt g1 till gK}
\end{proposition}
\begin{IEEEproof}
It is observed that \eqref{eq:robust constraint on SINR of Eve} differs from \eqref{eq:epigraph reformulation on gamma_e} in the only respect that \(\sigma_e^2\) is replaced by \(\tau\sigma_e^2\). Hence the proof for Proposition~\ref{prop:eqv LMI wrt g1 till gK} can be directly applied herein by substituting \(\tau\sigma_e^2\) for \(\sigma_e^2\).
\end{IEEEproof}

Last, by replacing ``\(\eta P_s\)'' in \eqref{eq:constraints on the jamming power}  with ``\(\tau\eta P_s\)'' in \eqref{eq:robust constraint on amount of AN for each HJ helper}, \eqref{eq:robust constraint on amount of AN for each HJ helper} can be replaced by a similar LMI as \eqref{eq:eqv LMI of available AN for S-Procedure}, denoted by $\rm (68e^\prime)$, in which the pertinent auxiliary variables are denoted by \(\{\mu_k\ge 0\}\).

Consequently, the equivalent reformulation for problem $\rm (P2^\prime.1\text{-}RW\text{-}SDR)$ can be summarized as
\begin{subequations}
\begin{align}
&\mathrm{(P2^\prime.1\text{-}RW\text{-}SDR)}: \mathop{\mathtt{max}}_{\boldsymbol{X},\{\boldsymbol{Q}_k\},\delta,\tau,\atop {s^{(0)}, s^{\prime(0)},s^{\prime\prime(0)},\atop \{s^{(k)}\},\{s^{\prime\prime(k)}\},\{\mu_k\}}}\delta\notag\\
&\mathtt{s.t.}~~~\eqref{eq:LMI of obj for S-Procedure}, \eqref{eq:LMI wrt tilde_h1 till tilde_hK}, \eqref{eq:LMI wrt g1 till gK}, {\rm (68e^\prime)}, \eqref{eq:robust constraint on transmit power of the relay}, \eqref{eq:robust constraint on PSD}, \notag\\
&\, s^{(0)}\ge 0, \, s^{\prime(0)}\ge 0, \, s^{\prime\prime(0)}\ge 0, \label{eq:non negative s0}\\
&s^{(k)}\ge 0, \, s^{\prime\prime(k)}\ge 0, \, \mu_k\ge 0, \; \forall k. \label{eq:non negative sk}
\end{align}
\end{subequations}

\subsection{Proposed Rank-One Solutions to $\rm (P2^\prime.1)$}\label{subsec:Proposed Solutions}
\(\mathrm{(P2^\prime.1\text{-}RW\text{-}SDR)}\) is convex and can be solved efficiently by convex optimization tools such as CVX. Next, we derive the Lagrangian of  \(\mathrm{(P2^\prime.1\text{-}RW\text{-}SDR)}\). \textcolor{black}{Note that in the following expression, we only consider the uncertainties regarding \(\tilde{\mv h}_0\), \(\mv h_k\)'s, \(\tilde{\mv h}_k\)'s, \(\mv g_0\) and \(\mv g_k\)'s when \(K=1\) for the purpose of simplicity and the results can be easily extended to the case of \(K>1\).} Denote the dual variables associated with \eqref{eq:robust constraint on transmit power of the relay}, \eqref{eq:LMI of obj for S-Procedure}, \eqref{eq:LMI wrt tilde_h1 till tilde_hK} and \eqref{eq:LMI wrt g1 till gK} by \(\beta_0\), \(\mv W\), \(\mv V\) and \(\mv Y\), respectively. \textcolor{black}{Then the partial Lagrangian of \(\mathrm{(P2^\prime.1\text{-}RW\text{-}SDR)}\) w.r.t.~\(\mv X\) is
\begin{align}
L(\overline{\mv\Omega})=\mathrm{tr}(\bar{\mv A}\mv X), \label{eq:Lagrangian of (P1-joint-Robust-SDR)}
\end{align}
where \(\overline{\mv\Omega}\) is the set of all primal and dual variables, and
\begin{equation}\label{eq:overline A}
\begin{aligned}
&\overline{\mv A}=P_s\mv W_{1,1}+2P_s\Re\{\hat{\mv f}_1^\dag\mv W_{12}^T\}+P_sw_{2,2}\hat{\mv F}_1\\
&-\sigma_r^2\sum_{i=1}^{N_t}\left(\mv V_{1,1}^{(i,i)}+2\Re\{\overline{\mv V}_{1,2}^{(i,i)}\}+\overline{\mv V}_{2,2}^{(i,i)}\right )\\
&-P_s(\mv h_0^H\otimes\mv I)\mv Y_{1,1}(\mv h_0\otimes\mv I)+2\sigma_r^2\bar\gamma_e\sum_{i=1}^{N_t^2}\Re\{\overline{\mv Y}_{2,1}^{(i,i)}\}\\
&-2P_s\Re\{(\mv h_0^H\otimes\mv I)\overline{\mv Y}_{2,1}(\mv h_0\otimes\mv I)\}\\
&-P_sy_{2,2}(\mv h_0^H\otimes\mv I)\hat{\mv g}^\dag\hat{\mv g}^T(\mv h_0\otimes\mv I)+\sigma_r^2\bar\gamma_e\sum_{i=1}^{N_t^2}\overline{\mv Y}_{2,2}^{(i,i)}.
\end{aligned}
\end{equation}
In \eqref{eq:overline A}, \(\hat{\mv F}_1=\hat{\mv f}_1^\dag\hat{\mv f}_1^T\); \(\mv W_{i,j}\), \(i,j=1,2\), \(\mv V_{i,j}\), \(i,j=1,\ldots,3\) and \(\mv Y_{i,j}\), \(i,j=1,\ldots,3\) are the block submatrices of \(\mv W\in\mathbb{C}^{(N_t^2+1)\times(N_t^2+1)}\), \(\mv V\in\mathbb{C}^{(N_t^3+N_t+1)\times(N_t^3+N_t+1)}\) and \(\mv Y\in\mathbb{C}^{(N_t^3+2N_t+1)\times(N_t^3+2N_t+1)}\) with the same size as block submatrices in \eqref{eq:LMI of obj for S-Procedure}, \eqref{eq:LMI wrt tilde_h1 till tilde_hK} and \eqref{eq:LMI wrt g1 till gK}, respectively.}
Moreover, in \eqref{eq:overline A}, \(\overline{\mv V}_{1,2}=\hat{\mv y}_1^\dag{\mv V}_{1,2}^T\), \(\overline{\mv V}_{2,2}= v_{2,2}\hat{\mv y}_1^\dag\hat{\mv y}_1^T\), \(\overline{\mv Y}_{2,1}=\hat{\mv g}^\dag\mv y_{1,2}^T\) and \(\overline{\mv Y}_{2,2}=y_{2,2}\hat{\mv g}^\dag\hat{\mv g}^T\). Furthermore, \(\mv V_{1,1}^{(i,i)}\), \(\overline{\mv V}_{1,2}^{(i,i)}\) and \(\overline{\mv V}_{2,2}^{(i,i)}\) are the \(i\)th block diagonal submatrices of \(\mv V_{1,1}\in\mathbb{C}^{N_t^3\times N_t^3}\), \(\overline{\mv V}_{1,2}\in\mathbb{C}^{N_t^3\times N_t^3}\) and \(\overline{\mv V}_{2,2}\in\mathbb{C}^{N_t^3\times N_t^3}\), respectively; \(\overline{\mv Y}_{2,1}^{(i,i)}\) and \(\overline{\mv Y}_{2,2}^{(i,i)}\) are the \(i\)th block diagonal submatrices of \(\mv Y_{2,1}\in\mathbb{C}^{N_t^3\times N_t^3}\), and \(\overline{\mv Y}_{2,2}\in\mathbb{C}^{N_t^3\times N_t^3}\), respectively.


\begin{proposition}
  \begin{enumerate}
  \item The optimal \(\mv X^\ast\) to $\rm (P2^\prime.1\text{-}RW\text{-}SDR)$ is expressed as
      \begin{align}
     \mv X^\ast=\sum_{n=1}^{N_t^2-\bar r_c}\bar a_n\bar{\mv \eta}_n\bar{\mv \eta}_n^H+\bar b\bar{\mv \xi}\bar{\mv \xi}^H, \label{eq:structure of robust X}
    \end{align}
    where \(\bar a_n\ge 0\), \(\forall n\),  \(\bar b>0\), and \(\bar{\mv\xi}\in\mathbb{C}^{N_t^2\times1}\) is a unit-norm vector orthogonal to \(\bar{\mv\Xi}\) (c.f.~\eqref{eq:structure of optimal X}).
    \item According to \eqref{eq:structure of robust X}, if \({\rm rank}(\mv X^\ast)>1\), i.e., there exists at least one \(\bar a_n>0\), we reconstruct a solution to problem $\rm (P2^\prime.1\text{-}RW\text{-}SDR)$ using
   \begin{align}
    \hat{\mv X}^\ast & =\bar b\bar{\mv\xi}\bar{\mv\xi}^H, \label{eq:reconstructed structure of suboptimal hat X}\\
    \hat\delta^\ast & =\delta^\ast, \label{eq:reconstructed suboptimal delta}\\
    \hat\tau^\ast & =\tau^\ast, \label{eq:reconstructed suboptimal tau}
    \end{align}
while \(\{\hat{\mv Q}_k^\ast\}\) are obtained by solving the following feasibility problem provided that \(\hat{\mv X}^\ast\), \(\hat\delta^\ast\), and \(\hat\tau^\ast\) are given by \eqref{eq:reconstructed structure of suboptimal hat X}, \eqref{eq:reconstructed suboptimal delta} and \eqref{eq:reconstructed suboptimal tau}, respectively:
    \begin{align*}
    &\mathrm{(P2^\prime.1\text{-}RW\text{-}SDR\text{-}sub)}:~ \mathop{\mathtt{max}}_{\{\boldsymbol{Q}_k\}, s^{\prime\prime(0)},\atop \{s^{\prime\prime(k)}\},\{\mu_k\}} 0\\
    &\mathtt{s.t.}~~~\eqref{eq:LMI wrt tilde_h1 till tilde_hK}\  \mbox{given}\ \hat{\mv X}^\ast, \hat\tau^\ast, \, \, {\rm (68e^\prime)}\ \mbox{given}\ \hat\tau^\ast,\\
    &\boldsymbol{Q}_k\succeq\mv 0,\, \mu_k\ge0, \; \forall k,\\
    &s^{\prime\prime(0)}\ge0,\, s^{\prime\prime(k)}\ge 0, \; \forall k.
    \end{align*}
  \end{enumerate}\label{prop:structure of robust X and its rank-one reconstruction}
\end{proposition}

\begin{IEEEproof}
See Appendix \ref{appendix:proof of prop:structure of robust X and its rank-one reconstruction}.
\end{IEEEproof}

The scheme that solves \(\mathrm{(P2^\prime)}\) is summarized in Table \ref{table:Algorithm II}.

\begin{table}[htp]
\begin{center}
\vspace{0.025cm}
\caption{\textcolor{black}{\rm Algorithm II for \(\mathrm{(P2^\prime)}\)}} \label{table:Algorithm II}
\vspace{-0.05cm}
 \hrule
\vspace{0.2cm}
\begin{itemize}
\item {\bf Initialize} \(\bar\gamma_{e\_{\rm search}}^\prime=0:\alpha^\prime:\bar\gamma_{e\max}^\prime\) and $i=0$
\item {\bf Repeat}
\begin{itemize}
\item [1)] {\bf Set} $i=i+1$;
\item [2)]  Given \(\bar\gamma_e=\bar\gamma_{e\_{\rm search}}^\prime(i)\),\\
 {\bf solve} \(\mathrm{(P2^\prime.1\text{-}RW\text{-}SDR)}\) and {\bf obtain} \(\hat H(\bar\gamma_e^{(i)})\).
\end{itemize}
\item {\bf Until} \(i=L^\prime\), where  \(L^\prime=\lfloor{\tfrac{\bar\gamma_{e\max}^\prime}{\alpha^\prime}}\rfloor+1\) is the length of \(\bar\gamma_{e\_{\rm search}}^\prime\)
\item {\bf Set} \(\bar\gamma_e^\ast=\bar\gamma_{e\_{\rm search}}^\prime\left(\!\arg\max\limits_{i}\!\left\{\tfrac{1+\hat H(\bar\gamma_e^{(i)})}{1+\bar\gamma_e^{(i)}}\right\}\right)\) for \(\mathrm{(P2^\prime.2)}\)
\item Given \(\bar\gamma_e^\ast\), {\bf solve} \(\mathrm{(P2^\prime.1\text{-}RW\text{-}SDR)}\) to obtain \((\mv X^\ast, \{\mv Q_k^\ast\}, \delta^\ast,\tau^\ast)\)\\
    {\bf if} \({\rm rank}(\mv X^\ast)=1\), {\bf apply} EVD on \(\mv X^\ast\) such that \(\mv X^\ast=\mv w^\ast\mv w^{\ast H}\);\\
    {\bf else}
    \begin{itemize}
     \item {\bf construct} \((\hat{\mv X}^\ast,\hat\delta^\ast,\hat\tau^\ast)\), according to \eqref{eq:reconstructed structure of suboptimal hat X}-\eqref{eq:reconstructed suboptimal tau} and {\bf set} \(\mv w^\ast=\sqrt{\bar b}\bar{\mv\xi}\);
     \item given \(\hat{\mv X}^\ast\), \(\hat\delta^\ast\) and \(\hat\tau^\ast\),\\
      {\bf obtain} \(\{\hat{\mv Q}_k^\ast\}\) by solving \(\mathrm{(P2^\prime.1\text{-}RW\text{-}SDR\text{-}sub)}\).
    \end{itemize}
    {\bf end}
    \item {\bf Recover} \(\mv W^\ast={\rm vec}^{-1}(\mv w^\ast)\)
\end{itemize}
\vspace{0.2cm} \hrule 
\end{center}
\end{table}

\section{Numerical Results}\label{sec:Numerical Results}
Here we provide numerical examples to validate our results. 
We assume a typical scenario where the \(K\) helpers are evenly distributed around Alice with a radius of \(\rho_k=2\)m and \(\theta_k=\tfrac{2\pi(k-1)}{K}\) (radian by default), where \(\theta_k\) is the angle of direction (w.r.t.~the Alice-relay link by default) of the \(k\)th helper, \(k=1,\ldots,K\). Alice, Bob and Eve are, w.l.o.g., assumed to have the same distance away from the AF relay with their angle of direction \(\pi\), \(\pi/6\) and \(11\pi/6\), respectively. We also assume channel models with both large-scale fading, i.e., path loss and shadowing, and small-scale fading, i.e., multi-path fading. The simplified large-scale fading model is given by \cite{wang2011general}
\begin{equation}\label{eq:shadowing+pass loss}
D=zA_0\left(\frac{d}{d_0}\right)^{-\alpha},~\mbox{for }d\geq d_0,
\end{equation}
\textcolor{black}{where \(z\) is a log-normal random variable capturing the effect of shadowing with the standard derivation \(\sigma=4\)dB, \(A_0=10^{-3}\), \(d\) is the distance, \(d_0\) is a reference distance set to be \(1\)m, and \(\alpha=2\) is the path loss exponent.} Specifically, the channels including \(\mv h_k\)'s, \(\mv h_0\), \(\tilde{\mv h}_0\) and \(\mv g_0\), are assumed to suffer from Rician fading while the channels from the  HJ helpers to Bob (\(\tilde{\mv h}_k\)'s) and Eve (\(\mv g_k\)'s) follow Rayleigh distribution due to the missing of line-of-sight (LOS) components with their respective average gain specified by \eqref{eq:shadowing+pass loss}. Take \(\mv h_k\), \(\forall k\), as an example, \(\mv h_k=\sqrt{\tfrac{K_R}{K_R+1}}\bar{\mv h}_k+\sqrt{\tfrac{1}{K_R+1}}\check{\mv h}_k\), where \(\bar{\mv h}_k\) is the LOS component with \(\|\bar{\mv h}_k\|_2^2=D\) (c.f.~\eqref{eq:shadowing+pass loss}), \(\check{\mv h}_k\) is the Rayleigh fading component denoted by \(\check{\mv h}_k\sim\mathcal{CN}(0,D\mv I)\), and \(K_R\) is the Rician factor set to be \(3\). Note that for the involved LOS component, we use the far-field uniform linear antenna array to model the channels \cite{karipidis2007far}.
In addition, unless otherwise specified, the number of HJ helpers, \(K\) is set to be \(5\); the AF relay is assumed to be \(5\)m away from Alice; the EH efficiency, \(\eta=0.5\) and \(\sigma_r^2=\sigma_b^2=\sigma_e^2=-50\)dBm. The results presented in Section \ref{subsec:The Perfect CSI Case} are obtained by averaging over \(500\) times of independent trials.

\subsection{The Perfect CSI Case}\label{subsec:The Perfect CSI Case}
We compare the proposed optimal solutions with three suboptimal schemes in the case of perfect CSI. \textcolor{black}{One suboptimal scheme, denoted by ``Suboptimal~1'', is introduced in Section~\ref{subsubsec:structure-based} by exploiting the optimal structure of \(\mv W\). The other  described in Section~\ref{subsubsec:ZF} is known as optimal null-space ZF, denoted by ``Suboptimal~2". Specifically, each jamming beam \(\mv n_k\) is restricted to lie in the orthogonal space of \(\tilde{\mv h}_k^\dag\) such that \(\mv n_k\)'s cause no interference to the IR while maximizing its effect of jamming at the eavesdropper. As a benchmark, we also present the well-known \emph{isotropic jamming} that is particularly useful when there is no Eve's CSI known at each \emph{HJ} helper, \({\sf H}_k\), \(\forall k\) \cite{Luo2013uncoordinated}, denoted by ``Suboptimal~3''. Note that the difference between ``Suboptimal~2'' and ``Suboptimal~3'' only lies in  the design of jamming noise, for which the former also aligns the jamming noise to an equivalent Eve's channel to confront Eve with most interference, while the latter transmits isotropic jamming with \(\tilde{\mv n}_k\sim\mathcal{CN}(\mv 0, \eta P_s\|\mv h_k\|^2/(N_t-1))\), \(k=1,\ldots, K\), in directions orthogonal to \(\tilde{\mv h}_k\)'s,  due to lack of knowledge of Eve's channel and thus is expected to be less efficient than ``Suboptimal~2'' with perfect CSI.}

\begin{figure}[tp]
\centering
 \epsfxsize=1\linewidth
   \includegraphics[width=7.5cm]{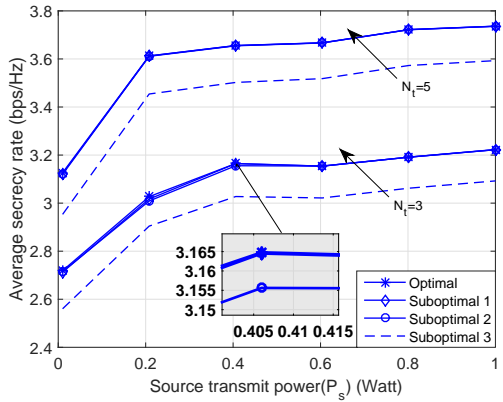}
\caption{{Secrecy rate versus Alice's transmit power with perfect CSI.}}\label{fig:sec_vs_Ps}
\end{figure}

First, we study the secrecy rate at the receiver versus the transmit power of the transmitter, \(P_s\) with \(P_r=10\)dBm. Fig.~\ref{fig:sec_vs_Ps} demonstrates that for both cases of \(N_t=3\) and \(N_t=5\), the average secrecy rate increases and tends to be saturated as \(P_s\) goes to \(30\)dBm. It also illustrates that ``suboptimal 1'' and ``suboptimal 2'' closely approach the optimal solutions while ``Suboptimal 3'' is outperformed more succinctly with larger number of antennas at the AF relay and the HJ helpers. Moreover, with \(N_t\) increasing, the average secrecy rate gets larger as a result of the higher array gain of the AF relay and more available power transferred to the HJ helpers.

\begin{figure}[tp]
\centering
 \epsfxsize=1\linewidth
   \includegraphics[width=7.5cm]{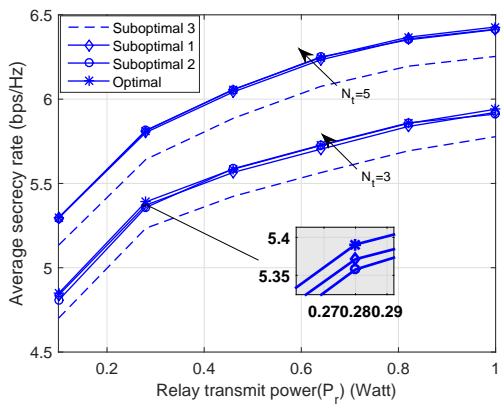}
\caption{{Secrecy rate versus the relay's transmit power with perfect CSI.}}\label{fig:sec_vs_Pr}
\vspace{-1.5em}
\end{figure}

In addition, we show in Fig.~\ref{fig:sec_vs_Pr} the secrecy rate achieved by different schemes versus the transmit power of the AF relay, \(P_r\) with \(P_s=30\)dBm. It is seen that the average secrecy rate first grows faster and then slower, since when \(P_r\) increases, not only the desired signal but also the noise yielded from the first transmission phase is amplified to a larger extent. In addition, the performance gap between the optimal scheme and suboptimal schemes is almost negligible. Similar to Fig.~\ref{fig:sec_vs_Ps}, \textcolor{black}{``Suboptimal 3'' appears to have certain performance loss from the optimality but is considered as a promising scheme when no Eve's CSI is available at the HJ helpers.}

\subsection{The Imperfect CSI Case}\label{subsec:The Imperfect CSI Case}
Now, we consider the imperfect CSI case and compare the proposed scheme \emph{Robust SDR with HJ}, which is obtained by solving $\rm (P2^\prime.1\text{-}RW\text{-}SDR\text{-}sub)$, against some benchmarks. \textcolor{black}{Note that there are two upper-bound benchmark schemes, namely, \emph{Robust SDR with HJ} and \emph{Robust-eqv with HJ}, as well as two lower-bound benchmarks, which are \emph{Robust w/o HJ} and \emph{Non-robust with HJ}. For \emph{Robust SDR with HJ} (\emph{Robust-eqv with HJ}), given any $\bar\gamma_e$, $\hat H(\bar\gamma_e)$ is approximated by solving the rank constraint relaxed problem $\rm (P2^\prime.1\text{-}RW\text{-}SDR)$ ($\rm (P2^\prime.1\text{-}RW\text{-}SDR\text{-}Eqv)$).} On the other hand, for \emph{Robust w/o HJ}, we solve $\rm (P2^\prime.1\text{-}RW\text{-}SDR)$  by setting $\mv Q_k=0$, $\forall k$ while for \emph{Non-robust with HJ}, \eqref{eq:achievable secrecy rate} is evaluated by applying the optimal solutions to $\rm (P1^\prime.1)$ assuming perfect CSI, to the actual channels including errors that are generated from the sets defined in \eqref{eq:uncertainty regions}.

To assess the worst-case secrecy performance, we use the metric, namely, \emph{secrecy outage probability}, defined as \cite{Liang2008}:
\begin{equation}\label{eq:sec outage prob}
p=P_r(r\le r_0^\ast),
\end{equation}
\textcolor{black}{where \(r_0^\ast\) obtained by solving $\rm (P2^\prime)$ is termed as the $100p\%$-\emph{secrecy outage rate}.}

The parameters are set identical to those in the perfect CSI case. Regarding the uncertainty model in \eqref{eq:uncertainty regions}, we introduce the uncertainty ratios associated with \(\epsilon_0\), \(\epsilon_0^\prime\), \(\epsilon_k\), \(\epsilon_k^\prime\) and \(\epsilon_k^{\prime\prime}\) as \(\alpha_0\), \(\alpha_0^\prime\), \(\alpha_k\), \(\alpha_k^\prime\) and \(\alpha_k^{\prime\prime}\), respectively. For instance, \(\alpha_0\) is
\begin{align}
\alpha_0^2 & =\frac{\epsilon_0}{\mathbb{E}[\|\boldsymbol{g}_0\|^2]},\label{eq:alpha0}
\end{align}
while \(\alpha_0^\prime\), \(\alpha_k\)'s, \(\alpha_k^\prime\)'s and \(\alpha_k^{\prime\prime}\)'s are similarly defined and thus omitted here for brevity. Besides, it is reasonable to assume that the channel estimates w.r.t~Eve suffer from more errors than those for Alice and Bob. Hence, we set \(\alpha_0^{\prime2}=\alpha_k^{\prime2}=\alpha_k^{\prime\prime2}=1\%\) while \(\alpha_0^2=\alpha_k^2=10\%\), \(k=1,\ldots,K\), unless otherwise specified.

\begin{figure}[tp]
\centering
 \epsfxsize=1\linewidth
   \includegraphics[width=7.5cm]{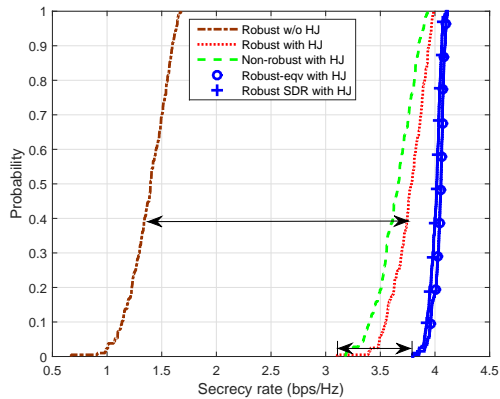}
\caption{\textcolor{black}{CDFs of the achievable secrecy rate.}}
\label{fig:CDF}
\vspace{-1.5em}
\end{figure}

Fig.~\ref{fig:CDF} demonstrates the cumulative density function (CDF) of the achievable secrecy rate from \(1000\) samples of random channel errors uniformly distributed over the sets defined by \eqref{eq:uncertainty regions} given fixed actual channel realization. We set \(P_r=20\)dBm, \(P_s=30\)dBm, \(N_t=3\), \(K=5\) and \(\alpha_0^{\prime2}=\alpha_k^{\prime2}=\alpha_k^{\prime\prime2}=2\%\), \(k=1,\ldots, K\). \textcolor{black}{Despite being suboptimal to the upper-bound schemes of ``Robust SDR with HJ'' and ``Robust-eqv with HJ'', the proposed ``Robust with HJ'' scheme outperforms its non-robust counterpart ``Non-robust with HJ'' particularly in the low range of probability, and overwhelmingly surpasses the ``Robust w/o HJ''. For example, ``Robust with HJ'' can achieve a secrecy rate of around \(3.5\)bps/Hz in the \(3\%\) worst case versus that of \(3.3\)bps/Hz and \(1.0\)bps/Hz for the ``Non-robust with HJ'' and ``Robust w/o HJ'', respectively. The solutions for ``Robust SDR with HJ'' is also seen to admit very little gap from those for ``Robust-eqv with HJ'', which suggests that approximating \(\hat H(\bar\gamma_e)\) by solving the complexity reduced ``Robust SDR with HJ '' leads almost no performance loss.}

\begin{figure}[tp]
\centering
 \epsfxsize=1\linewidth
   \includegraphics[width=7.5cm]{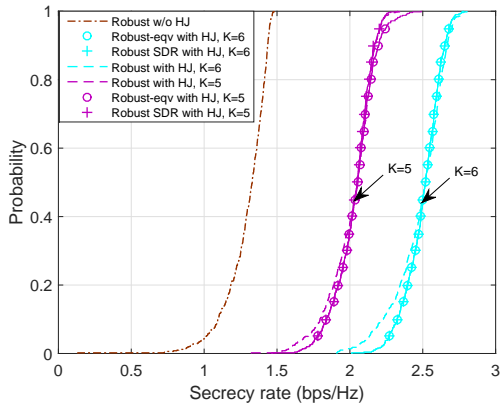}
\caption{{Secrecy outage probability for \(K=3\) and \(K=5\) HJ helpers, respectively.}} \label{fig:outage_vs_K}
\vspace{-1.5em}
\end{figure}

Fig.~\ref{fig:outage_vs_K} illustrates the CDF of the achievable secrecy rate from \(1000\) samples of random channel errors generated in the same way as Fig.~\ref{fig:CDF}, with \(P_r=20\)dBm, \(P_s=30\)dBm and \(N_t=3\). It is observed that proposed solutions to ``Robust with HJ'' nearly achieve their upper-bound rank constraint relaxed solutions, i.e., SDR, to ``Robust upper SDR with HJ'' throughout the whole range of outage probability. Moreover, the ``Robust w/o HJ'' yields the worst performance. In particular, when the outage probability falls to \(3\%\), the ``Robust w/o HJ'' achieves a worst-case secrecy rate of less than \(1\)bps/Hz while the proposed scheme can still guarantee an outage rate of rough \(1.64\)bps/Hz and \(2.07\)bps/Hz for \(K=5\) and \(K=6\), respectively. \textcolor{black}{Also, it is observed that increasing the number of HJ helpers will improve the secrecy performance, but we do not draw conclusions on the extent to which the secrecy rate can increase, since it also depends on the level of channel estimation inaccuracy. For example, more HJ helpers may also yield larger interference to the legitimate receiver if the channels from HJ helpers to Bob are not as well estimated as this instance of \(\alpha_k^{\prime\prime2}=1\%\), \(\forall k\).} Hence we suggest that in practice, a mild number of HJ helpers are sufficient in view of the trade-off between complexity and performance.

\begin{figure}[tp]
\centering
 \epsfxsize=1\linewidth
   \includegraphics[width=7.5cm]{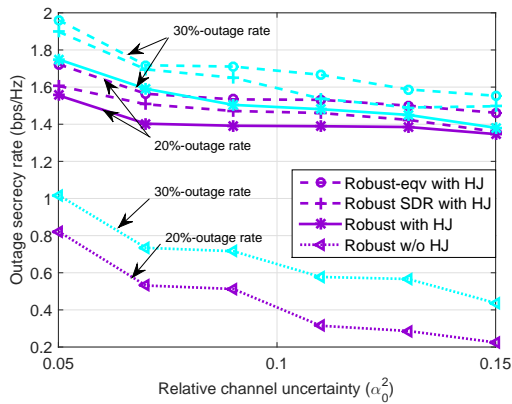}
\caption{{Secrecy outage rate versus the normalized channel errors.}} \label{fig:outage_vs_error}
\vspace{-1.5em}
\end{figure}

Fig.~\ref{fig:outage_vs_error} shows two different levels (\(p=0.20\) and \(p=0.30\)) of secrecy outage rate versus the channel uncertainty ratios (assuming \(\alpha_0=\alpha_k\), \(k=1,\ldots,K\)), in which \(P_r=30\)dBm, \(P_s=30\)dBm, \(N_t=3\) and \(K=5\). It is observed that the secrecy outage rate by the proposed schemes decreases slowly with the eavesdropper's CSI error ratios, which validates the motivation of the worst-case robust optimization. It is worth noting that the advantage of the \emph{HJ} protocol is more significant when the normalized channel uncertainty of Eve's channels  surpasses \(10\%\), since the \emph{HJ} scheme provides more degree of freedom for robust design and thus capable of guaranteeing larger worst-case secrecy rate against worse channel conditions compared to that without \emph{HJ}. The reasonably suboptimal performance of the proposed ``Robust with HJ'' is also seen as from Figs.~\ref{fig:CDF} and \ref{fig:outage_vs_K}.

\begin{figure}[tp]
\centering
 \epsfxsize=1\linewidth
   \includegraphics[width=7.5cm]{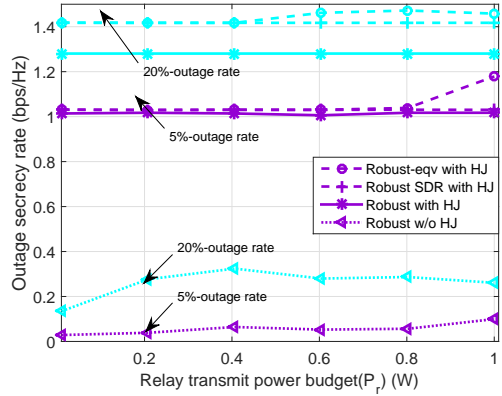}
\caption{{Secrecy outage rate versus the relay's transmit power.}} \label{fig:outage_vs_Pr}
\vspace{-1.5em}
\end{figure}

Fig.~\ref{fig:outage_vs_Pr} studies the \(100p\%\)-secrecy outage rate for \(p=0.05\) and \(p=0.20\), respectively, versus the transmit power of the AF relay. Specifically, we set \(P_s=30\)dBm, \(N_t=3\), and \(K=5\). As observed similarly from Fig.~\ref{fig:outage_vs_error}, the robust schemes with the assistance of HJ helpers perform considerably better than solutions without HJ helpers. \textcolor{black}{Furthermore, when the transmit power is set relatively large, i.e., \(P_s=30\)dBm, it is seen that  continuously increasing \(P_r\) does not contribute much to the secrecy performance, because in this situation the increased amplified noise at the AF relay compromises the performance, which provides useful insight for practical setting of \(P_r\). In addition, the proposed ``Robust with HJ'' is observed striking a good trade-off between optimality and complexity compared with the two upper-bound solutions.}

\section{Conclusion}\label{sec:Conclusion}
This paper considered improving the secret wireless communications in a multi-antenna AF relay wiretap channel via a novel {\em harvest-and-jam (HJ)} relaying protocol. 
The AN covariance matrices at HJ helpers and the AF relay beamforming matrix have been jointly optimized to maximize the achievable secrecy rate and/or worst-case secrecy rate at the legitimate receiver subject to the transmit power constraints of the AF relay as well as the HJ helpers, on perfect and imperfect CSI occasions, respectively, using the technique of semi-definite relaxation (SDR). The SDR was shown tight for the perfect CSI case while suboptimal rank-one reconstruction algorithms for the robust formulation under imperfect CSIs were presented achieving promising tradeoffs between complexity and performance. The effectiveness of the proposed schemes were also verified by numerical results.

\begin{appendix}

\subsection{Proof of Proposition \ref{prop:structure of optimal X and its rank-one reconstruction}} \label{appendix:proof of prop:structure of optimal X and its rank-one reconstruction}
The KKT conditions of $\rm (P1^\prime.1\text{-}RW\text{-}SDR)$ are given by
\begin{subequations}\label{eq:KKT of (P1'.1-joint-SDR)}
\begin{align}
\mv A^\ast\mv X^\ast &=0, \label{eq:complimentary slackness wrt X}\\
\mv B_k^\ast\mv Q_k^\ast&=0, \forall k, \label{eq:complimentary slackness wrt Qk}\\
\beta_k^\ast\left(\mathrm{tr}(\mv Q_k^\ast)-\tau^\ast\eta P_s\|\mv h_k\|^2\right)&=0, \forall k. \label{eq:complimentary slackness wrt trace(Qk)}
\end{align}
\end{subequations}
According to \eqref{eq:Bk}, if for certain \(k\), \(\beta_k^\ast=0\), then \(\mv B_k^\ast=-\lambda^\ast\tilde{\mv h}_k^\ast\tilde{\mv h}_k^T+\alpha^\ast\bar\gamma_e\mv g_k^\ast\mv g_k^T\) and thus \(\mathrm{rank}(\mv B_k^\ast)\le \mathrm{rank}(\tilde{\mv h}_k^\ast\tilde{\mv h}_k^T)+\mathrm{rank}(\mv g_k^\ast\mv g_k^T)=2\), which yields \(\mathrm{rank}(\mv Q_k^\ast)\ge N_t-2\) as a result of \eqref{eq:complimentary slackness wrt Qk}. Otherwise, when \(\beta_k^\ast>0\), we will have \(\mathrm{rank}(\mv B_k^\ast)\ge\mathrm{rank}(-\beta_k^\ast\mv I-\lambda^\ast\tilde{\mv h}_k^\ast\tilde{\mv h}_k^T)-\mathrm{rank}(\alpha^\ast\bar\gamma_e\mv g_k^\ast\mv g_k^T)=N_t-1\) \cite[\emph{Lemma~A.1}]{Liu2014Secrecy}, which implies \(\mathrm{rank}(\mv Q_k^\ast)\le 1\). However, \(\mathrm{rank}(\mv Q_k^\ast)\) cannot be \(0\),  since otherwise \(\mathrm{tr}(\mv Q_k^\ast)-\tau^\ast\eta P_s\|\mv h_k\|^2<0\) and thus \(\beta_k^\ast=0\) according to \eqref{eq:complimentary slackness wrt trace(Qk)}, which contradicts to \(\beta_k^\ast>0\). Hence, when \(\beta_k^\ast>0\), \(\mathrm{rank}(\mv Q_k^\ast)=1\).

Next, define \(\mv C^\ast=-\lambda^\ast\sigma_r^2\overline{\mv Y}_1-\alpha^\ast P_s\mv F_2+\alpha^\ast\bar\gamma_e\sigma_r^2\overline{\mv Y}_2-\beta_0^\ast\overline{\mv \Phi}\) and according to \eqref{eq:A}, we have
\begin{align}\label{eq:A star}
\mv A^\ast=P_s\mv F_1+\mv C^\ast.
\end{align}
Then define \(r_c\), \(\mv\Xi\) and \(\mv\eta_n\), \(n=1,\ldots,N_t^2-r_c\) (c.f.~\eqref{eq:structure of optimal X}). Similar to the approach used in \cite[Appendix B]{Liu2014Secrecy}, we discuss the structure of the optimal \(\mv X\) under two cases.

\begin{enumerate}[(1)]
  \item {\bf Case I}: \(r_c=N_t^2\).
  As \(\mv C^\ast\) is full-rank, \(\mathrm{rank}(\mv A^\ast)\ge r_c-1=N_t^2-1\) and hence \(N_t^2-1\le\mathrm{rank}(\mv A^\ast)\le N_t^2\). If \(\mathrm{rank}(\mv A^\ast)=N_t^2-1\), \(\mathrm{rank}(\mathbf{null}(\mv A^\ast))=1\) and  it follows that \(\mv X^\ast=b\mv\xi\mv\xi^H\) by assuming \(\mv\xi\) as the only  basis of \({\bf null}(\mv A^\ast)\). Otherwise, according to \eqref{eq:complimentary slackness wrt X}, we obtain \(\mv X^\ast=\mv 0\), which ceases the secrecy transmission and  cannot be the optimal solution to $\rm (P1^\prime.1\text{-}RW\text{-}SDR)$.

\item{\bf Case II}: \(r_c<N_t^2\).
If \(\mv C^\ast\) is not full-rank, \(\mathrm{rank}(\mv A^\ast)\ge r_c-1\). Then by pre-multiplying \(\mv\eta_n^H\) and post-multiplying \(\mv\eta_n\in\mv\Xi\) with both sides of \eqref{eq:A star}, we have
\begin{align}\label{eq:null(C star) in null(A star)}
\mv\eta_n^H\mv A^\ast\mv\eta_n=P_s\mv\eta_n^H\mv F_1\mv\eta_n+\mv\eta_n^H\mv C^\ast\mv\eta_n=P_s\mv\eta_n^H\mv F_1\mv\eta_n,\; \forall n.
\end{align}
According to \eqref{eq:Lagrangian of (P1.1-RW-SDR)}, it is necessary for \(\mv A^\ast\preceq\mv 0\) to obtain an optimal solution of \(\mv X^\ast\) and therefore \(\mv\eta_n^H\mv A^\ast\mv\eta_n\le 0\), which conforms to \(P_s\mv\eta_n^H\mv F_1\mv\eta_n\ge 0\) if and only if \(\mv A^\ast\mv\eta_n=0\) and \(\mv F_1\mv\eta_n=0\). Hence, \(\mv\Xi\subseteq{\bf null}(\mv A^\ast)\), i.e., \(N_t^2-\mathrm{rank}(\mv A^\ast)\ge N_t^2-r_c\Rightarrow\mathrm{rank}(\mv A^\ast)\le r_c\). 
Next, we show \(\mathrm{rank}(\mv A^\ast)\neq r_c\) by contradiction. If \(\mathrm{rank}(\mv A^\ast)=r_c\), \(\mv\Xi={\bf null}(\mv A^\ast)\), and \(\mv X^\ast=\sum_{n=1}^{N_t^2-r_c}a_n\mv\eta_n\mv\eta_n^H\). However, in this case, since \(\mv F_1\mv\eta_n=0\), \(P_s\mathrm{tr}(\mv F_1\mv X^\ast)=0\), which is apparently not optimal. Hence, we have \(\mathrm{rank}(\mv A^\ast)=r_c-1\) and thus \(\mathrm{rank}({\bf null}(\mv A^\ast))=N_t^2-r_c+1\). This indicates that besides the basis in \(\mv\Xi\), \({\bf null}(\mv A^\ast)\) spans over an extra dimension of basis, which is denoted by \(\mv\xi\), and hence \(\mv X^\ast=\sum_{n=1}^{N_t^2-r_c}a_n\mv\eta_n\mv\eta_n^H+b\mv\xi\mv\xi^H\).
\end{enumerate}

Assume that \((\mv X^\ast,\{\mv Q_k^\ast\},\tau^\ast)\) is the  optimal solution to $\rm (P1^\prime.1\text{-}RW\text{-}SDR)$ with \(\mathrm{rank}(\mv X^\ast)>1\). Then construct a new solution \(\{\hat{\mv X}^\ast,\hat{\mv Q}_k^\ast,\hat\tau^\ast\}\) according to \eqref{eq:reconstructed structure of optimal hat X}--\eqref{eq:reconstructed tau}. Now, we check if the reconstructed solution is feasible if \eqref{eq:sufficient condition for rank-one X} holds.
First,
\begin{align}
&\sigma^2_r{\rm tr}(\overline {\mv Y}_1\hat{\mv X}^\ast)+\sum_{k=1}^K\tilde{\mv h}_k^T\hat{\mv Q}_k^\ast\tilde{\mv h}_k^\dagger+\hat\tau^\ast\sigma_b^2\notag\\
&\le\sigma^2_r{\rm tr}\left(\overline {\mv Y}_1\left(\mv X^\ast-\sum_{n=1}^{N_t^2-r_c}a_n\mv\eta_n\mv\eta_n^H\right)\right)+\sum_{k=1}^K\tilde{\mv h}_k^T\mv Q_k^\ast\tilde{\mv h}_k^\dagger\nonumber\\
&+\left(\tau^\ast+\tfrac{\sigma_r^2}{\sigma_b^2}\sum_{n=1}^{N_t^2-r_c}a_n\mathrm{tr}(\overline{\mv Y}_1\mv\eta_n\mv\eta_n^H)\right)\sigma^2_b\notag\\
&=\sigma^2_r{\rm tr}(\overline {\mv Y}_1\mv X^\ast)+\sum_{k=1}^K\tilde{\mv h}_k^T\mv Q_k^\ast\tilde{\mv h}_k^\dagger+\tau^\ast\sigma^2_b\stackrel{(a)}{\le}1.\label{eq:reconstructed C-O transformation constraint}
\end{align}
Moreover,
\begin{align}
&P_s{\rm tr}(\mv F_2\hat{\mv X}^\ast)=P_s{\rm tr}\left(\mv F_2(\mv X^\ast-\sum_{n=1}^{N_t^2-r_c}a_n\mv\eta_n\mv\eta_n^H)\right)\nonumber\\
&\stackrel{(b)}{\le}\bar\gamma_e\left(\sigma^2_r{\rm tr}(\overline {\mv Y}_2\mv X^\ast)+\sum_{k=1}^K\mv g_k^T\mv Q_k^\ast\mv g_k^\dagger+\tau^\ast\sigma_e^2\right)\nonumber\\
&+\bar\gamma_e\left(\sigma^2_e\Delta\tau-\sigma^2_r{\rm tr}\left(\overline{\mv Y}_2\sum_{n=1}^{N_t^2-r_c}a_n\mv\eta_n\mv\eta_n^H\right)\right)\nonumber\\
&=\bar\gamma_e\left(\sigma^2_r{\rm tr}(\overline {\mv Y}_2\hat{\mv X}^\ast)+ \sum_{k=1}^K\mv g_k^T\hat{\mv Q}_k^\ast\mv g_k^\dagger+\hat\tau^\ast\sigma^2_e\right).\label{eq:reconstructed SINR of Eve constraint}
\end{align}
In addition, \eqref{eq:constraint on transmit power of the relay}--\eqref{eq:constraint on X, Qk and tau} are easily shown to satisfy.
%
In the above, \((a)\) and \( (b)\) hold due to the feasibility in \eqref{eq:constraint on C-O transformation} and \eqref{eq:constraint on SINR of Eve}, respectively. Further, \(P_s{\rm tr}(\mv F_1\hat{\mv X}^\ast)=P_s\mathrm{tr}(\mv F_1\mv X^\ast)\)  shows that the reconstructed solution achieves the same optimum value as that of $\rm (P1^\prime.1\text{-}RW\text{-}SDR)$. Hence, an optimal solution to $\rm (P1^\prime.1\text{-}RW\text{-}SDR)$ with rank-one \(\mv X\) is ensured.

\subsection{Proof of Lemma \ref{lemma:optimal structure for W}} \label{appendix:proof of lemma:optimal structure for W}
First, we construct \(\mv W\) as
\begin{align}\label{eq:decomposition of W}
 \mv W & =[\mv U_1^\dag,(\mv U_1^{\bot})^\dag]\begin{bmatrix}
\mv B & \mv C \\
 \mv D& \mv E
\end{bmatrix}[\mv U_2,\mv U_2^{\bot}]^H\notag\\
& =\mv U_1^\dag\mv B\mv U_2^H+\mv U_1^\dag\mv C(\mv U_2^{\bot})^H+(\mv U_1^{\bot})^\dag\mv D\mv U_2^H\nonumber\\
&+(\mv U_1^{\bot})^\dag\mv E(\mv U_2^{\bot})^H,
\end{align}
where \(\mv B\in\mathbb{C}^{2\times2}\), \(\mv C\in\mathbb{C}^{2\times (N_t-2)}\), \(\mv D\in\mathbb{C}^{(N_t-2)\times 2}\) and \(\mv E\in\mathbb{C}^{(N_t-2)\times (N_t-2)}\) are undetermined matrices. Then according to \eqref{eq:SVD of H1} and \eqref{eq:SVD of H2}, it follows that \(\vert\widetilde{\mv{h}}^T_0\mv{Wh}_0\vert^2=\vert\widetilde{\mv h}_0^T\mv U_1^\dag\mv B\mv U_2^H\mv h_0\vert^2\) and \(\widetilde{\mv{h}}^T_0\mv{WW}^H\widetilde{\mv{h}}_0^\dagger=\|\mv B^H\mv U_1^T\widetilde{\mv h}_0^\dagger\|^2+\|\mv C^H\mv U_1^T\widetilde{\mv h}_0^\dagger\|^2\). Similarly, we also have \(\vert{\mv{g}}^T_0\mv{Wh}_0\vert^2=\vert{\mv g}_0^T\mv U_1^\dag\mv B\mv U_2^H\mv h_0\vert^2\) and \({\mv{g}}^T_0\mv{WW}^H\mv g_0^\dagger=\|\mv g_0^T\mv U_1^\dagger\mv B\|^2+\|\mv g_0^T\mv U_1^\dagger\mv C\|^2\). Thus, \(\gamma_b\) (c.f.~\eqref{eq:SINR at Bob}) and \(\gamma_e\) (c.f.~\eqref{eq:SINR at Eve}) do not depend on \(\mv D\) and \(\mv E\).

Next, by substituting \eqref{eq:decomposition of W} for \(\mv W\) in \eqref{eq:transmit power at the relay}, we have \(P_r\ge P_s(\|\mv B\mv U_2^H\mv h_0\|^2+\|\mv D\mv U_2^H\mv h_0\|^2)+\sigma_r^2\mathrm{tr}(\mv B^H\mv B+\mv C^H\mv C+\mv D^H\mv D+\mv E^H\mv E )\).
 Since $\rm (P1^\prime)$ is a secrecy rate maximization problem subject to the given \(P_r\), it turns out that given the optimum secrecy rate, \(P_r\) is the minimized required power by taking \(\mv D=\mv 0\) and \(\mv E=\mv 0\), while \(\mv B\) and \(\mv C\) cannot be determined directly. Thus, \(\mv W=\mv U_1^\dagger\mv B\mv U_2^H+\mv U_1^\dagger\mv C(\mv U_2^{\bot})^H\).

\subsection{Proof of Proposition \ref{prop:semi-closed form}}\label{appendix:proof of prop:semi-closed form}
Denoting the dual variable associated with \eqref{eq:C1 of ZF}, \eqref{eq:C2 of ZF} and \eqref{eq:C3 of ZF} by \(\lambda\), \(\alpha\) and \(\beta_0\), respectively, the Lagrangian of \(\mathrm{(P1^\prime.1\text{-}sub2\text{-}SDR)}\) is expressed as
\begin{multline}
  L(\chi)=\\
  {\rm tr}\left((P_s\mv F_1-\lambda\sigma_r^2\overline{\mv Y}_1-\alpha P_s\mv F_2+\alpha\bar\gamma_e\sigma_r^2\overline{\mv Y}_2-\beta_0\overline{\mv\Phi})\mv X \right )\\
  +\left(-\lambda\sigma_b^2+\alpha\bar\gamma_e(q+\sigma_e^2)+\beta_0P_r\right )\tau+\lambda, \label{eq:Lagrangian of (P1'.1-sub2-SDR)}
\end{multline}
where \(\chi=\{\mv X,\tau,\lambda,\alpha,\beta_0\}\) denotes the set consisting of all the primal and dual variables. Since problem \(\mathrm{(P1^\prime.1\text{-}sub2\text{-}SDR)}\) satisfies the Slater condition, its optimum value admits zero duality gap with its dual counterpart. Furthermore, according to \eqref{eq:Lagrangian of (P1'.1-sub2-SDR)}, in order for the dual function to be bounded from above, the following constraints must hold:
\begin{align}
&\!\!\!\!\!\!\mv Z= P_s\mv F_1-\lambda\sigma_r^2\overline{\mv Y}_1-\alpha P_s\mv F_2+\alpha\bar\gamma_e\sigma_r^2\overline{\mv Y}_2-\beta_0\overline{\mv\Phi}\preceq\mv 0, \label{eq:bounded Lagrangian for Z}\\
&\!\!\!\!-\lambda\sigma_b^2+\alpha\bar\gamma_e(q+\sigma_e^2)+\beta_0P_r\le 0. \label{eq:bounded Lagrangian for tau}
\end{align}
The dual problem is therefore given by
\begin{subequations}
\begin{align}
\mathrm{(D\text{-}P1^\prime.1\text{-}sub2\text{-}SDR)}:\!\!\mathop{\mathtt{min}}_{\lambda,\alpha,\beta_0}\!\!&~\lambda\nonumber\\
\mathtt{s.t.}&~\eqref{eq:bounded Lagrangian for Z}, \eqref{eq:bounded Lagrangian for tau},\\
&~(\lambda,\alpha,\beta_0)^T\ge \mv 0.
\end{align}
\end{subequations}
It is observed that \(\mv Z\) is of the same form as the Hessian matrix with respect to \(\mv X\) without rank relaxation. According to \cite[Theorem~2.1]{Ai2009},  \(\mv Z\preceq\mv 0\) implies that the SDR problem \(\mathrm{(P1^\prime.1\text{-}sub2\text{-}SDR)}\) is tight in this case, i.e., \(\exists\mv w^\ast\) such that $\mv X^\ast=\mv w^\ast\mv w^{\ast H}$. Moreover, since KKT condition necessitates \(\mv Z^\ast\mv X^\ast=\mv 0\), it follows that \(\mv w^\ast\) is the eigenvector corresponds to the zero-eigenvalue of \(\mv Z^\ast\). Hence, we have \(\mv w^\ast=\mu\nu_{\max}(\mv Z^\ast)\), where \(\mu=\sqrt{\tfrac{P_r}{{\rm tr}(\overline{\mv\Phi})\mv\nu_{\max}(\mv Z^\ast)\mv\nu_{\max}^H(\mv Z^\ast)}}\) is due to the power constraint of \eqref{eq:constraint on transmit power of the relay}, which completes the proof.

\subsection{Proof of Proposition \ref{prop:eqv LMI wrt tilde_h1 till tilde_hK}}\label{appendix:proof of prop:eqv LMI wrt tilde_h1 till tilde_hk}
First, given \(\tilde{\mv h}_k\), \(k=2,\ldots,K\), fixed, only consider the uncertainty of \(\tilde{\mv h}_1\). Since \(\|\Delta\tilde{\mv h}_1\|_2^2\le\epsilon_1^{\prime\prime}\), we have \(1-\frac{(\Delta\tilde{\boldsymbol{h}}_1^\dag)^H\Delta\tilde{\boldsymbol{h}}_1^\dag}{\epsilon_1^{\prime\prime}}\ge 0\). By applying Lemma \ref{lemma:LMI from robust block QMI} to \eqref{eq:LMI of eqv obj for S-Procedure} with \({\mv H_1}^{(1)}=P_s\mv X^{\prime\prime}-\delta\sigma_r^2\mv X^\prime+w^{(0)}\mv I\), \({\mv F_1}^{(1)}=(P_s\mv X^{\prime\prime}-\delta\sigma_r^2\mv X^\prime)\hat{\tilde{\mv h}}^\dag\), \({\mv G_1}^{(1)}=\mv 0\),
$c_1^{(1)}=\hat{\tilde{\mv h}}^T(P_s\mv X^{\prime\prime}-\delta\sigma_r^2\mv X^\prime)\hat{\tilde{\mv h}}^\dag-\delta\hat{\tilde{\mv h}}_1^T\mv Q_1\hat{\tilde{\mv h}}_1^\dag-\delta\sum_{i=2}^K{\tilde{\mv h}}_i^T\mv Q_i{\tilde{\mv h}}_i^\dag-\delta\sigma_b^2-w^{(0)}N_t\epsilon_0^\prime$, \({\mv B_1}^{(1)}=-\delta\mv Q_1\hat{\tilde{\mv h}}_1^\dag\), and \({\mv A_1}^{(1)}=-\delta\mv Q_1\), there exists \(w^{(1)}\ge 0\) such that the following LMI holds:{\small
\begin{equation}\label{eq:LMI of eqv tilde h1 for S-Procedure}
\begin{bmatrix} {\mv H}_1^{(1)}&{\mv F_1}^{(1)}&{\mv G_1}^{(1)}\\
{\mv F_1}^{(1)H}&  c_1^{(1)} &{\mv B_1}^{(1)H}\\
{\mv G_1}^{(1)H}& {\mv B_1}^{(1)}&{\mv A_1}^{(1)}\end{bmatrix}-w^{(1)}\begin{bmatrix} \mv 0&\mv 0&\mv 0\\
\mv 0 & { 1} & \mv 0\\
\mv 0 & \mv 0 & \frac{-\mv I}{\epsilon_1^{\prime\prime}}\end{bmatrix}\succeq{\mv 0}.
\end{equation}}
Note that for \(\mv Q_1\succeq{\mv 0}\), there always exists \(w^{(1)}>0\) such that \(\tfrac{w^{(1)}\mv I}{\epsilon_1^{\prime\prime}}+{\mv A_1}^{(1)}\succ\mv 0\) and we assume that such constraint is applied. According to the property of Schur-Complements \cite[A.~5.5]{boyd2004convex}, for \eqref{eq:LMI of eqv tilde h1 for S-Procedure}, we have {\small
\begin{equation}\label{eq:array of eqv tilde_h1 for S-procedure}
\left\{\begin{aligned}
&\begin{bmatrix}
{\mv H_1}^{(1)}&{\mv F_1}^{(1)}\\
{\mv F_1}^{(1)H}& c_1^{(1)}-w^{(1)}\end{bmatrix}\\
&-\begin{bmatrix} {\mv G_1}^{(1)}\\ {\mv B_1}^{(1)H}\end{bmatrix}\left({\mv A_1}^{(1)}+\tfrac{w^{(1)}\mv I}{\epsilon_1^{\prime\prime}} \right )^{-1}\begin{bmatrix}{\mv G_1}^{(1)H}&{\mv B_1}^{(1)} \end{bmatrix}\succeq\mv 0,\\
&\tfrac{w^{(1)}\mv I}{\epsilon_1^{\prime\prime}}+{\mv A_1}^{(1)}\succ \mv 0,
\end{aligned}\right.
\end{equation}}
which can be reexpressed as {\small
\begin{align}
\begin{bmatrix} {\mv A_1}^{(1)}+\tfrac{w^{(1)}\mv I}{\epsilon_1^{\prime\prime}}&{\mv G_1}^{(1)H}&{\mv B_1}^{(1)}\\
{\mv G_1}^{(1)}& {\mv H_1}^{(1)} &{\mv F_1}^{(1)}\\
{\mv B_1}^{(1)H}&{\mv F_1}^{(1)H}&c_1^{(1)}-w^{(1)}\end{bmatrix}\succeq\mv 0. \label{eq:immediate LMI of eqv tilde_h1 for iteration}
\end{align}}

Next, assume that the robust design for \eqref{eq:LMI of eqv obj for S-Procedure} has been considered against the precedent \(k-1\) uncertainties, i.e.,{\small
\begin{align}
 &\begin{bmatrix} {\mv H_1}^{(k-1)}&{\mv F_1}^{(k-1)}&{\mv G_1}^{(k-1)}\\
{\mv F_1}^{(k-1)H}&  c_1^{(k-1)} &{\mv B_1}^{(k-1)H}\\
{\mv G_1}^{(k-1)H}& {\mv B_1}^{(k-1)}&{\mv A_1}^{(k-1)}\end{bmatrix}
-w^{(k-1)}\begin{bmatrix} \mv 0&\mv 0&\mv 0\\
\mv 0 & 1 & \mv 0\\
\mv 0 & \mv 0 &  \frac{-\mv I}{\epsilon_{k-1}^{\prime\prime}}\end{bmatrix}\nonumber\\
&\succeq\mv 0, \; k\ge2. \label{eq:LMI of eqv (k-1)s uncertainties for S-procedure}
\end{align}}
Applying a similar procedure as that for \eqref{eq:LMI of eqv tilde h1 for S-Procedure}, \eqref{eq:LMI of eqv (k-1)s uncertainties for S-procedure} can be recast as {\small
\begin{equation}\label{eq:immediate LMI of eqv (k-1)s uncertainties for iteration}
\begin{bmatrix} \tfrac{w^{(k-1)}\mv I}{\epsilon_{k-1}^{\prime\prime}}+{\mv A_1}^{(k-1)}&{\mv G_1}^{(k-1)H}&{\mv B_1}^{(k-1)}\\
{\mv G_1}^{(k-1)}& {\mv H_1}^{(k-1)} &{\mv F_1}^{(k-1)}\\
{\mv B_1}^{(k-1)H}& {\mv F_1}^{(k-1)H}&c_1^{(k-1)}-w^{(k-1)}\end{bmatrix}\succeq\mv 0.
\end{equation}}
Then given \(\tilde{\mv h}_i\), \(i=k+1,\ldots,K\) fixed, accommodate the \(k\)th uncertainty, i.e., \(\tilde{\mv h}_k\in\tilde{\mathcal H}_k\), for \eqref{eq:immediate LMI of eqv (k-1)s uncertainties for iteration}. By applying Lemma \ref{lemma:LMI from robust block QMI} to the uncertainty of \(\tilde{\mv h}_k\), the implication \( \|\Delta\tilde{\mv h}_k\|_2^2\le\epsilon_k^{\prime\prime}\Rightarrow\eqref{eq:immediate LMI of eqv (k-1)s uncertainties for iteration}\) holds if and only if there exists \(w^{(k)}\ge 0\) such that {\small
\begin{equation}
 \begin{bmatrix}
{\mv H_1}^{(k)}&{\mv F_1}^{(k)} &{\mv G_1}^{(k)} \\
{\mv F_1}^{(k)H} &  c_1^{(k)}  & {\mv B_1}^{(k)H} \\
{\mv G_1}^{(k)H}& {\mv B_1}^{(k)}& {\mv A_1}^{(k)}
\end{bmatrix}-w^{(k)}\begin{bmatrix} \mv 0 & \mv 0 &\mv 0\\\mv 0 &1 &\mv 0\\ \mv 0&\mv 0 &\frac{-\mv I}{\epsilon_{k}^{\prime\prime}} \end{bmatrix}\succeq {\mv 0},
\end{equation}}
where
{\small\begin{align}
&{\mv H_1}^{(k)}=\begin{bmatrix}{\mv A_1}^{(k-1)}+\frac{w^{(k-1)}\mv I}{\epsilon_{k-1}^{\prime\prime}}& {\mv G_1}^{(k-1)H}\\ {\mv G_1}^{(k-1)}& {\mv H_1}^{(k-1)} \end{bmatrix},\nonumber\\
&{\mv F_1}^{(k)}=\begin{bmatrix}{\mv B_1}^{(k-1)}\\ {\mv F_1}^{(k-1)} \end{bmatrix},\
{\mv G_1}^{(k)}=\mv 0,
\end{align}}
$c_1^{(k)}=\hat{\tilde{\mv h}}^T(P_s\mv X^{\prime\prime}-\delta\sigma_r^2\mv X^\prime)\hat{\tilde{\mv h}}^\dag-\delta\sum_{j=1}^k\hat{\tilde{\mv h}}_j^T\mv Q_j\hat{\tilde{\mv h}}_j^\dag-\delta\sum_{i=k+1}^K{\tilde{\mv h}}_i^T\mv Q_i{\tilde{\mv h}}_i^\dag-\delta\sigma_b^2-w^{(0)}N_t\epsilon_0^\prime-\sum_{l=1}^{k-1}w^{(l)}$, ${\mv B_1}^{(k)}=-\delta\mv Q_k\hat{\tilde{\mv h}}_k^\dag$ and ${\mv A_1}^{(k)}=-\delta\mv Q_k$, $k\ge 2$. Thus, using the method of mathematical induction, \eqref{eq:LMI of eqv obj for S-Procedure} holds for \(\tilde{\mv h}_k\in\tilde{\mathcal{H}}_k\), \(k=1,\ldots, K\),  if and only if there exists \(\{w(k)\ge 0\}\), such that \eqref{eq:LMI reformulation on delta} is satisfied, which completes the proof.

\subsection{Proof of Proposition \ref{prop:eqv LMI wrt g1 till gK}}\label{appendix:proof of prop:eqv LMI wrt g1 till gk}
Taking the similar procedure as that for dealing with \eqref{eq:semi-indefinite form of tilde_h for S-procedure},  the implication \(\|\Delta{\mv g}\|^2\le N_t\epsilon_0\Rightarrow\eqref{eq:linear epigraph reformulation on gamma_e}\) holds if and only if there exists \(v^{(0)}\ge 0\) such that the following LMI holds:{\small
\begin{align}
\begin{bmatrix}\mv H_2 & \mv F_2\\
\mv F_2^H & c_2 \end{bmatrix}\succeq\mv 0, \label{eq:LMI of eqv Eve's SINR for S-Procedure}
\end{align}}
where \(\mv H_2=-P_s\mv X^{\prime\prime}+\bar\gamma_e\sigma_r^2\mv X^\prime+v^{(0)}\mv I\), \(\mv F_2=(-P_s\mv X^{\prime\prime}+\bar\gamma_e\sigma_r^2\mv X^\prime)\hat{{\mv g}}^\dag\) and \(c_2=\hat{{\mv g}}^T(-P_s\mv X^{\prime\prime}+\bar\gamma_e\sigma_r^2\mv X^\prime)\hat{{\mv g}}^\dag+\bar\gamma_e\sum_{k=1}^K{{\mv g}}_k^T\mv Q_k{{\mv g}}_k^\dag+\bar\gamma_e^2-v^{(0)}N_t\epsilon_0\). \eqref{eq:epigraph reformulation on gamma_e} has been equivalently reformulated into \eqref{eq:LMI of eqv Eve's SINR for S-Procedure}. Then, given \({\mv g}_k\), \(k=2,\ldots,K\), fixed, applying similar procedure to that in Appendix~\ref{appendix:proof of prop:eqv LMI wrt tilde_h1 till tilde_hk}, it follows that there exists \(v^{(1)}\ge 0\) such that the following LMI holds:{\small
\begin{equation}\label{eq:LMI of eqv g1 for S-Procedure}
\begin{bmatrix} {\mv H}_2^{(1)}&{\mv F_2}^{(1)}&{\mv G_2}^{(1)}\\
{\mv F_2}^{(1)H}&  c_2^{(1)} &{\mv B_2}^{(1)H}\\
{\mv G_2}^{(1)H}& {\mv B_2}^{(1)}&{\mv A_2}^{(1)}\end{bmatrix}-v^{(1)}\begin{bmatrix} \mv 0&\mv 0&\mv 0\\
\mv 0 & { 1} & \mv 0\\
\mv 0 & \mv 0 & \frac{-\mv I}{\epsilon_1}\end{bmatrix}\succeq{\mv 0}.
\end{equation}}
Since \(\tfrac{v^{(1)}\mv I}{\epsilon_1}+{\mv A_2}^{(1)}\succ {\mv 0}\) always holds, \eqref{eq:LMI of eqv g1 for S-Procedure} is equivalent to the following LMI:
{\small
\begin{align}
\begin{bmatrix} {\mv A_2}^{(1)}+\tfrac{v^{(1)}\mv I}{\epsilon_1} &{\mv G_2}^{(1)H}&{\mv B_2}^{(1)}\\
{\mv G_2}^{(1)}& {\mv H_2}^{(1)} &{\mv F_2}^{(1)}\\
{\mv B_2}^{(1)H}&{\mv F_2}^{(1)H}&c_2^{(1)}-v^{(1)}\end{bmatrix}\succeq\mv 0. \label{eq:immediate LMI of eqv g1 for iteration}
\end{align}}
Next, devising the method of mathematical induction again as that for \eqref{eq:immediate LMI of eqv tilde_h1 for iteration}, \eqref{eq:LMI of eqv Eve's SINR for S-Procedure} holds for \(\mv g_k\in\mathcal{G}_k\), \(\forall k\), if and only if there exists \(\{v(k)\ge 0\}\), such that \eqref{eq:LMI reformulation on gamma_e} is satisfied, which completes the proof.

\subsection{Proof of Proposition \ref{prop:LMI wrt tilde_h1 till tilde_hK}}\label{appendix:proof of prop:LMI wrt tilde_h1 till tilde_hK}
We only sketch the proof herein since it is quite similar to that of Proposition~\ref{prop:eqv LMI wrt tilde_h1 till tilde_hK}. First, apply Lemma~\ref{lemma:LMI from robust block QMI} to \eqref{eq:LMI of y1 for S-Procedure} given \(\tilde{\mv h}_k\)'s, \(k=2,\ldots,K\), fixed and obtain an initial LMI. Next, manipulate the resulting LMI according to the property of Schur-Complements to facilitate using Lemma~\ref{lemma:LMI from robust block QMI}. Then, repeat this procedure until all the semi-indefinite constraints w.r.t.~\(\tilde{\mv h}_k\)'s have been incorporated into an equivalent LMI.

\subsection{Proof of Proposition \ref{prop:structure of robust X and its rank-one reconstruction}}\label{appendix:proof of prop:structure of robust X and its rank-one reconstruction}
According to the KKT conditions of $\rm (P2^\prime.1\text{-}RW\text{-}SDR)$, we have \(\bar{\mv A}^\ast\mv X^\ast={\bf 0}\), where \(\bar{\mv A}^\ast\) is given by \eqref{eq:overline A}. Define \(\bar{\mv C}^\ast=\bar{\mv A}^\ast-w_{2,2}^\ast P_s\hat{\mv F}_1\) with \({\rm rank}(\bar{\mv C}^\ast)\) denoted by \(\bar r_c\).
Then take the similar procedure as {\bf Case I} and {\bf Case II} in Appendix \ref{appendix:proof of prop:structure of optimal X and its rank-one reconstruction}, it can be obtained that \(\mv X^\ast=\sum_{n=1}^{N_t^2-\bar r_c}\bar a_n\bar{\mv \eta}_n\bar{\mv \eta}_n^H+\bar b\bar{\mv \xi}\bar{\mv \xi}^H\).

Next, we prove the second half of Proposition \ref{prop:structure of robust X and its rank-one reconstruction}. According to \eqref{eq:reconstructed structure of suboptimal hat X},
\begin{align}\label{eq:reconstructed robust constraint on obj}
\!\! P_s{\rm tr}(\hat{\mv F}_1\hat{\mv X}^\ast)
 =P_s{\rm tr}(\hat{\mv F}_1\mv X^\ast)
\ge\!\!\min\limits_{\tilde{\boldsymbol{h}}_0\in\tilde{\mathcal H}_0}\!\!P_s{\rm tr}(\mv F_1\mv X)\ge\delta^\ast,\!\!
\end{align}
 and thus \eqref{eq:robust constraint on delta} holds true, which implies that the same optimal value as $\rm (P2^\prime.1\text{-}RW\text{-}SDR)$, i.e., \(\delta^\ast\), is achievable. However, since the constraint in \eqref{eq:robust constraint on SINR of Eve} is ignored, the global optimal \(\bar\gamma_e^\ast\) for $\rm (P2^\prime.2)$ via solving $\rm (P2^\prime.1\text{-}RW\text{-}SDR)$ is probably violated in $\rm (P2^\prime.1\text{-}RW\text{-}SDR\text{-}sub)$. For example, \(\tfrac{P_s{\rm tr}(\boldsymbol{F}_2\hat{\boldsymbol{X}}^\ast)}{\sigma^2_r{\rm tr}(\overline {\boldsymbol{Y}}_2\hat{\boldsymbol{X}}^\ast)+\sum_{k=1}^K\boldsymbol{g}_k^T\hat{\boldsymbol{Q}}_k^\ast\boldsymbol{g}_k^\dagger+\hat\tau^\ast\sigma_e^2}=\bar\gamma_e^{0}\ge\hat F(\bar\gamma_e^\ast)\), which results in the actual objective value for $\rm (P2^\prime.2)$, \(\tfrac{1+\hat H(\bar\gamma_e^\ast)}{1+\bar\gamma_e^{0}}\) smaller than \(\tfrac{1+\hat H(\bar\gamma_e^\ast)}{1+\hat F(\bar\gamma_e^\ast)}\), and thus suboptimal for $\rm (P2^\prime)$.

\end{appendix}

\bibliographystyle{IEEEtran}
\balance
\bibliography{GC2014_jv}

\end{document}